\documentclass[a4paper,twocolumn,BCOR1.6cm,DIV10,numbers=noenddot]{scrartcl}

\usepackage{abstract}
\usepackage{amsbsy}
\usepackage{amsfonts}
\usepackage{amsmath}
\usepackage{amssymb}
\usepackage{array}
\usepackage{bbding}
\usepackage{bbm}
\usepackage{bm}
\usepackage{booktabs}
\usepackage[british]{babel}%
\usepackage{calc}
\usepackage[hang,small,bf]{caption}
\usepackage{cite}
\usepackage{color}
\usepackage{epsfig}
\usepackage{extarrows}
\usepackage{fmtcount}%http://dound.com/2009/06/latex-adding-footnotes-in-tables-or-other-floats/
\usepackage{footmisc}
\usepackage[T1]{fontenc}
\usepackage{graphicx}
\usepackage{german}
\usepackage[latin1]{inputenc}
\usepackage{longtable}
\usepackage{lscape}
\usepackage{mathcomp}
\usepackage{mathrsfs}
\usepackage{multirow}
\usepackage{needspace}
\usepackage[section]{placeins}
\usepackage{psfrag}
\usepackage{subfigure}
\usepackage{textcomp}
\usepackage{upgreek}
\usepackage{verbatim}
\usepackage{wasysym}
\usepackage{flushend}

\usepackage{wrapfig}
\usepackage{xr}
\usepackage{yfonts}
\usepackage{caption}
\usepackage[svgnames]{xcolor}
\usepackage{tikz}
\usetikzlibrary{calc,shapes}
\usepackage{ulem}

\usepackage{scrpage2}
\usepackage{times}

\usepackage{widetext}
\usepackage{balance}

\DeclareCaptionFont{white}{\color{white}}
\DeclareCaptionFont{black}{\color{black}}

\DeclareMathAlphabet\mathbfcal{OMS}{cmsy}{b}{n}

\allowdisplaybreaks[1]
\newcommand{\bls}{1.1}
\renewcommand{\baselinestretch}{\bls}

\providecommand*{\I}{\mathrm{i}}                           %% imaginary unit i
\providecommand*{\ket}[1]{|#1\rangle}                      %% ket vector
\newcommand{\rket}[1]{|#1)}                           %% ket with parenthesis
\renewcommand{\vec}[1]{\mathbf{#1}}                       %% vectors are bold symbols here
\newcommand{\nn}{\nonumber}

\renewcommand{\d}{\mathrm{d}}

\newcommand{\ketp}[1]{\left|\right.\hspace{-0.5ex}{#1}\left.\hspace{-0.5ex}\right)}

\newcommand{\lbrap}[1]{\left(\right.\hspace{-0.5ex}\widetilde{#1}\left.\hspace{-0.5ex}\right|}
\newcommand{\lbokp}[3]{\left(\right.\hspace{-0.5ex}\widetilde{#1}\left.\hspace{-0.5ex}\right|{#2}\left|\right.\hspace{-0.5ex}{#3}\left.\hspace{-0.5ex}\right)}

\newcommand{\pgftextcircled}[2][0]{
        \setbox0=\hbox{#2}%
        \dimen0\wd0%
        \divide\dimen0 by 2%
        \begin{tikzpicture}[baseline=(a.base)]%
            \useasboundingbox (-\the\dimen0,0pt) rectangle (\the\dimen0,1pt);
            \node[
                circle,
                draw,
                outer sep=0pt,
                inner sep=0.1ex
            ] (a) {#2};
        \end{tikzpicture}
    }

\definecolor{transpa}{rgb}{0.01, 0.0, 0.0}

\definecolor{darkgreen}{rgb}{0.0, 0.7, 0.3}
\newcommand{\calBvec}{\mathbfcal{B}}
\newcommand{\calEvec}{\mathbfcal{E}}

\usepackage{etoolbox}
\newtoggle{Nebenrechnung}
\togglefalse{Nebenrechnung} % comment out for version including all side calculations

\newtoggle{paragraph}
\toggletrue{paragraph} % comment out for no paragraph titles
\newtoggle{ToDo}
\togglefalse{ToDo} % comment out for version including all side calculations

\renewcommand{\thesubsection}{\arabic{subsection}}

\newcommand{\ie}{i.~e.}
\newcommand{\eg}{e.~g.}
\newcommand{\wrt}{with respect to~}
\newcommand{\9}{{9{;}\hspace{-0.05em}11}}
\newcommand{\1}{{11\hspace{-0.05em}{;}9}}

\newcommand{\smmath}[1]{\mbox{$#1$}}

\begin{document}

\title{\vspace{-4em}\Large{Geometric phases causing lifetime modifications\\ of metastable states of hydrogen}}

\author{\large M.-I.~Trappe,$^{1,2,}$\footnotemark[1]\hspace{0.75ex} %$^,$\footnotemark[2]\hspace{1ex},
 \large P.~Augenstein,$^{3,}$\footnotemark[2]\hspace{1ex}
 \large M.~DeKieviet,$^{3,}$\footnotemark[3]\hspace{1ex}
 \large T.~Gasenzer,$^{2,4,5,}$\footnotemark[4]\hspace{1ex}
 \large O.~Nachtmann$^{2,}$\footnotemark[5]\\
 \footnotesize
 $^{1}${\it Centre for Quantum Technologies, National University of Singapore, 3 Science Drive 2, Singapore 117543, Singapore}\\[-0.6em]
 \footnotesize
 $^{2}${\it Institut f\"ur Theoretische Physik, Universit\"at Heidelberg, Philosophenweg 16, 69120 Heidelberg, Germany}\\[-0.6em]
 \footnotesize
 $^{3}${\it Physikalisches Institut, Universit\"at Heidelberg, Im Neuenheimer Feld 226, 69120 Heidelberg, Germany}\\[-0.6em]
 \footnotesize
 $^{4}${\it Kirchhoff-Institut f\"ur Physik, Universit\"at Heidelberg, Im Neuenheimer Feld 227, 69120 Heidelberg, Germany}\\[-0.6em]
 \footnotesize
 $^{5}${\it ExtreMe Matter Institute EMMI, GSI Helmholtzzentrum f{\"u}r Schwerionenforschung, Planckstra{\ss}e 1, 64291 Darmstadt, Germany}}

\date{\small(\today)}

\renewcommand{\abstractname}{}
\twocolumn[
  \maketitle
  \begin{onecolabstract}
    \vspace{-6em}
    {%\begin{flushleft}
    \noindent
    Externally applied electromagnetic fields in general have an influence on the width of atomic spectral lines.
    The decay rates of atomic states can also be affected by the geometry of an applied field configuration giving rise to an imaginary geometric phase.
    A specific chiral electromagnetic field configuration is presented which geometrically modifies the lifetimes of metastable states of hydrogen. 
    We propose to extract the relevant observables in a realistic longitudinal atomic beam spin-echo apparatus which allows the initial and final fluxes of the metastable atoms to be compared with each other interferometrically. 
    A geometry-induced change in lifetimes at the $5\%$-level is found, an effect large enough to be observed in an available experiment.\newline
    }
    \vspace{2em}
  \end{onecolabstract}
]

\small

%==========================================================================
%==========================================================================
\subsection{Introduction}{\renewcommand{\thefootnote}{}\footnote{\begin{flushleft}\vspace{-1.5\baselineskip}$^*$martin.trappe@quantumlah.org\\ $^\dagger$augenstein@physi.uni-heidelberg.de\\$^\ddag$maarten.dekieviet@physik.uni-heidelberg.de\\$^\S$t.gasenzer@uni-heidelberg.de\\$^\P$O.Nachtmann@thphys.uni-heidelberg.de\end{flushleft}}}%
Atoms being exposed to an adiabatically varying external field can acquire geometric phases \cite{Ber84,Bar83}. 
For metastable states, such geometric phases are in general complex.
The imaginary part of such a phase influences the lifetime, see e.g.~\cite{Garrison1988177,Miniatura1990,Massar1996,Berry2004}.

In Refs.~\cite{BeGaNa07_I,BeGaNa07_II,BeGaMaNaTr08_I,DeKGaNaTr11,GaNaTr2012}, we have presented studies of geometric phases for metastable states of hydrogen.
Both, parity-conserving (PC) and parity-violating (PV) geometric phases were identified. 
It was, in particular, shown in \cite{GaNaTr2012} that the lifetimes of metastable 2S hydrogen states can be influenced by geometric phases acquired by the atom in suitable external electric and magnetic fields. 
A concrete example of the influence of a complex geometric phase on the lifetime of atomic states was discussed in \cite{GaNaTr2012}. 
With the field configurations investigated there geometric effects on the lifetimes at the per mille level were found. 

In the present paper we shall explore suitable field configurations which lead, in theory, to geometric effects on the lifetimes of metastable hydrogen states up to the level of several per cent.
We propose to measure the lifetime shifts by means of an existing longitudinal atomic beam spin-echo interferometer that allows the initial and final fluxes of metastable atoms to be compared with each other. The results presented here were obtained by means of the theoretical formalism introduced in detail in Refs.~\cite{BeGaMaNaTr08_I,GaNaTr2012}. 
We refer to these papers for the discussion of the general context of our investigations and of the proposed experimental scheme, as well as for many further references. 
We will, in particular, make use of specific expressions and formulae from these papers, referring to them without repeating their derivations.

%==========================================================================
%==========================================================================
\subsection{Metastable hydrogen in the longitudinal atomic beam spin-echo apparatus}\label{SectionMetastable}
\captionsetup[figure]{format=plain,font={scriptsize,normalfont}}
\subsubsection{Atomic-beam spin-echo interferometer}
%=====================================
\renewcommand{\baselinestretch}{1.0}
\begin{figure}[h!tbp]
\centering
\includegraphics[width=0.99\linewidth]{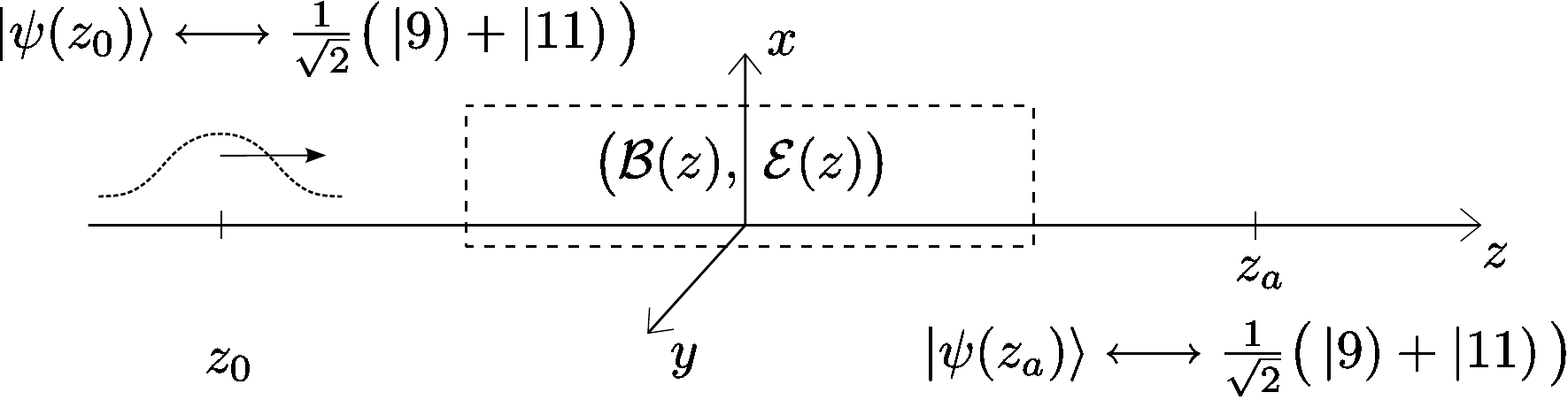}
\caption{Scheme of the atom interferometry experiment. The atom is prepared around $z_0$ and analysed around $z_a$. We start with a superposition $\ket{\psi(z_0)}$ of the two states $\ketp{9}$ and $\ketp{11}$. After passing the electric and magnetic fields 
%$\calEvec(z)$ and $\calBvec(z)$, 
the wave function is projected onto an analysing state $\ket{\psi(z_a)}$, for example, again onto a superposition of the states $\ketp{9}$ and $\ketp{11}$. The coordinate axes used, $x$, $y$, $z$, indexed in the formulae as 1, 2, and 3, respectively, are also indicated.}
\label{SchematiclABSE}
\end{figure}
\renewcommand{\baselinestretch}{\bls}
\clearcaptionsetup{figure}
%=====================================
As in \cite{BeGaMaNaTr08_I} we consider metastable 2S hydrogen states in the spin-echo interferometer described in \cite{ABSE95}.\linebreak[4] {Figure~\ref{SchematiclABSE}} shows a schematic view of the atomic-beam spin-echo interferometer. An atomic state, in general a superposition of local energy eigenstates, enters the interferometer at $z_0$. The state is then subjected to electric and magnetic fields 
$\calEvec(z)$ and $\calBvec(z)$. 
Finally, it is analysed at $z_a$ by projection on a chosen final state. In reference to the experiment, we set in the following
\begin{align}
\begin{split}
z_0&=0\,\mathrm{m}\ ,\\
z_a&=0.66\,\mathrm{m}\ .
\end{split}
\end{align}
First we consider field configurations of a general type, consisting of two regions I and II in space and/or time of the spin-echo setup \cite{ABSE95}, in which the spins precess forward and backwards, respectively (thus separated by an effective $\pi$-pulse). These regions have an electric field
\begin{align}
\calEvec(z)&=\calEvec_{\mathrm{I}}(z)\Theta(\smmath{\frac12}z_a-z)\Theta(z)\nn\\
&\phantom=+\calEvec_{\mathrm{II}}(z)\Theta(z_a-z)\Theta(z-\smmath{\frac12}z_a)\ ,\label{E}
\end{align}
and a magnetic field with the components
\begin{align}
\calBvec(s;z)&=\vec{e}_1\mathcal B_1(z)+\vec{e}_2\mathcal B_2(z)+\vec{e}_3\mathcal B_3(s;z)\ ,\label{B}
\end{align}
where
\begin{align}
\mathcal B_{i}(z)&=\mathcal B_{i\mathrm{I}}(z)\Theta(\smmath{\frac12}z_a-z)\Theta(z)\nn\\
&\phantom=+\mathcal B_{i\mathrm{II}}(z)\Theta(z_a-z)\Theta(z-\smmath{\frac12}z_a)
\end{align}
for $i=1,2$, and
\begin{align}\label{B3si}
\mathcal B_3(s;z)&=\mathcal B_{3\mathrm{I}}(z)\Theta(\smmath{\frac12}z_a-z)\Theta(z)\nn\\
&\phantom=+s\,\mathcal B_{3\mathrm{II}}(z)\Theta(z_a-z)\Theta(z-\smmath{\frac12}z_a)\ .
\end{align}
We also require
\begin{align}
\begin{split}\label{zerofields}
\calEvec(0)&=\calEvec(\smmath{\frac12}z_a)=\calEvec(z_a)=0\ ,\\
\calBvec(s;0)&=\calBvec(s;\smmath{\frac12}z_a)=\calBvec(s;z_a)=0\ .
\end{split}
\end{align}
In (\ref{E})-(\ref{B3si}) $\Theta(\cdot)$ is the usual step function and $s$ is a parameter, which acts as a detuning between the spin precession regions I and II, and is varied around the spin echo point, $s=1$, by typically
\begin{align}
0.4\le s\le1.6\ .
\end{align}
The variation of $s$, that is, the variation of the magnetic field $\mathcal B_3$ in the second half of the interferometer produces the oscillations in the spin-echo signal; see \cite{BeGaMaNaTr08_I}. Explicit examples of external fields within this general form are given in Section \ref{SectionExamples} below (see Figures \ref{example1fields}--\ref{FieldConfigSymmCond}).

An atom travelling through the interferometer with field configuration (\ref{E})-(\ref{zerofields}) traces out, in parameter space, a closed path $C_s$, where $s$ is kept fixed. In fact, $C_s$ is composed of two successive paths in regions I and II,
\begin{align}\label{path}
C_s=C_{\mathrm{I}}+C_{\mathrm{II},s}\ .
\end{align}
We shall now consider field configurations that, in parameter space, correspond to oppositely oriented paths, either along the reverse of the complete path $C$, or along the reverse of the paths $C_{\mathrm{I}}$ and $C_{\mathrm{II}}$ separately.

For reversing the complete path $C$ we consider the fields
\begin{align}\label{rev}
\begin{split}
\calEvec^{\mathrm{rev}}(z)&=\calEvec(z_a-z)\ ,\\
\mathcal B^{\mathrm{rev}}_i(z)&=\mathcal B_i(z_a-z)\ ,\quad
\mbox{for }i=1,2\ ,\\
\mathcal B^{\mathrm{rev}}_{3}(s;z)&=s\,\mathcal B_{3\mathrm{II}}(z_a-z)\Theta(\smmath{\frac12}z_a-z)\Theta(z)\\
&\quad+\mathcal B_{3\mathrm{I}}(z_a-z)\Theta(z_a-z)\Theta(z-\smmath{\frac12}z_a)\ .
\end{split}
\end{align}
From (\ref{E})--(\ref{zerofields}) and (\ref{rev}) we see that, in the reverse field configuration, the atomic system traces out the path which is the reversed one of (\ref{path}),
\begin{align}
\overline C_s=\overline C_{\mathrm{II},s}+\overline C_{\mathrm{I}}\ .
\end{align}
Note that for the reverse field configuration the magnetic field component $\mathcal B_3$ is varied with $s$ in the first half of the interferometer.

For the second case of reversing the paths in regions I and II of the interferometer separately, we consider the following fields:
\begin{align}\label{rev2}
\begin{split}
\tilde{\calEvec}^{\mathrm{rev}}(z)&=\calEvec_{\mathrm{I}}(\smmath{\frac12}z_a-z)\Theta(\smmath{\frac12}z_a-z)\Theta(z)\\
&\phantom=+\calEvec_{\mathrm{II}}(\smmath{\frac32}z_a-z)\Theta(z_a-z)\Theta(z-\smmath{\frac12}z_a)\ ,\\
\tilde{\mathcal B}^{\mathrm{rev}}_i(z)&=\mathcal B_{i\mathrm{I}}(\smmath{\frac12}z_a-z)\Theta(\smmath{\frac12}z_a-z)\Theta(z)\\
&\phantom=+\mathcal B_{i\mathrm{II}}(\smmath{\frac32}z_a-z)\Theta(z_a-z)\Theta(z-\smmath{\frac12}z_a)\ ,\\
\mbox{for }i&=1,2\ ,\\
\tilde{\mathcal B}^{\mathrm{rev}}_3(s;z)&=\mathcal B_{3\mathrm{I}}(\smmath{\frac12}z_a-z)\Theta(\smmath{\frac12}z_a-z)\Theta(z)\\
&\phantom=+s\,\mathcal B_{3\mathrm{II}}(\smmath{\frac32}z_a-z)\Theta(z_a-z)\Theta(z-\smmath{\frac12}z_a)\ .
\end{split}
\end{align}
Here the path of the atom in parameter space in relation to (\ref{path}) is
\begin{align}
\overline C'_s=\overline C_{\mathrm{I}}+\overline C_{\mathrm{II},s}\ .
\end{align}

\subsubsection{Hydrogen spin-echo observables}
The hydrogen states under investigation are 2S states that are admixed with 2P states in external electric fields. Our numbering of the 16 ($n=2$)-states of hydrogen is explained in detail in Appendix A, Table A.2, of \cite{GaNaTr2012}. The index set of metastable states is
\begin{align}
I=\{9,10,11,12\}\ .
\end{align}

The initial state at $z=z_0$ is a superposition of metastable states
\begin{align}\label{calpha}
\begin{split}
\left.\ket{\psi(0)}\right|_{\mathrm{internal}}&=\sum_{\alpha\in I}c_\alpha\ketp{\alpha(z_0)}\ ,\\
\sum_{\alpha\in I}|c_\alpha|^2&=1\ .
\end{split}
\end{align}
See (72) in \cite{BeGaMaNaTr08_I} for the complete state vector. Here and in the following we write out only the internal part of it. In (\ref{calpha}) and in the following $\ketp{\alpha(z)}$ ($\alpha=1,\dots,16$) are the local energy right eigenstates corresponding to the fields $\calEvec(z)$, $\calBvec(z)$; see (13) of \cite{BeGaMaNaTr08_I}. 

As discussed in \cite{BeGaMaNaTr08_I}, the effective potentials $\mathcal V_\alpha(z)$ entering the Schr\"odinger equation for the atomic states in the external fields are not equal to the local complex energy eigenvalues $E_\alpha(z)$, see (31)--(33) of \cite{BeGaMaNaTr08_I}, as they include additional geometric-phase effects. But, as we shall show below, in our case this difference is negligible. Nonetheless, we work in the following with the effective potentials as this is the correct procedure. The value of the effective potential for the state $\alpha$ at point $z$ is in general complex
\begin{align}
\mathcal V_\alpha(z)=\mathrm{Re}\,\mathcal V_\alpha(z)-\frac{\I}{2}\Gamma_\alpha(z)\ .
\end{align}
Here
\begin{align}
\Gamma_\alpha(z)=-2\mathrm{Im}\,\mathcal V_\alpha(z)
\end{align}
is the local decay rate of the state $\alpha$; see (32), (33) of \cite{BeGaMaNaTr08_I}. For the field configurations considered in the present work, we find for $\alpha=9,11$
\begin{align}\label{ReVE}
|\mathrm{Re}\,\big(\mathcal V_\alpha(z)-E_\alpha(z)\big)|\lesssim 10^{-16}\,\mathrm{eV}\ ,
\end{align}
and
\begin{align}\label{ImVE}
\frac{|\mathrm{Im}\,\big(\mathcal V_\alpha(z)-E_\alpha(z)\big)|}{\underset{0\le z\le z_a}{\mathrm{max}}|\mathrm{Im}\,E_\alpha(z)|}\lesssim 10^{-10}\ ,
\end{align}
that is, the numerical differences between $\mathcal V_\alpha(z)$ und $E_\alpha(z)$ are negligible since we shall deal with energies at the $\mu$eV scale; cf. Figure \ref{example1energies} below. 

The atoms in the beam have typical longitudinal velocity $v_z$, wave number $k_z$ and de Broglie wavelength $\lambda$ (see (20) of \cite{BeGaMaNaTr08_I})
\begin{align}
v_z&=\frac{k_z}{m}\approx 3500 \,\mathrm{m/s}\ ,\nn\\
k_z&\approx5.6\times10^{10}\,\mathrm m^{-1}\ ,\nn\\
\lambda&=\frac{2\pi}{k_z}\approx 1.1\times 10^{-10}\,\mathrm m\ .
\end{align}
At the end of the interferometer, at $z=z_a$, the atomic state is projected onto a chosen state (see (90) of \cite{BeGaMaNaTr08_I})
\begin{align}\label{palpha}
\begin{split}
\ketp{p}&=\sum_{\alpha\in I}p_\alpha\ketp{\alpha(0)}\ ,\\
\sum_{\alpha\in I}|p_\alpha|^2&=1\ .
\end{split}
\end{align}
The integrated flux $\mathcal F_p$ for this state is the experimental observable 
\begin{align}\label{Fp}
{\mathcal F}_p=\sum_{\alpha,\beta\in I}&p_\beta p^*_\alpha c^*_\beta c_\alpha
\exp [-(\Delta\tau_\beta-\Delta\tau_\alpha)^2/(8\sigma'^2_k)]\nn\\
&\quad\times U^*_\beta(z_a,z_0;\bar k_m)\,U_\alpha(z_a,z_0;\bar k_m)\ .
\end{align}
All quantities occuring in (\ref{Fp}) are defined and explained in the context of Eq.~(105) in \cite{BeGaMaNaTr08_I}. 
We briefly recall them in the following.

The $U_\alpha$ contain the dynamic and geometric phases, see (101) of \cite{BeGaMaNaTr08_I},
\begin{align}\label{Ualpha}
U_\alpha(z_a,z_0;\bar k_m)=
\exp[-\I\varphi_\alpha(z_a)+\I\gamma_\alpha(z_a)]\,.
\end{align}
Here $\bar k_m$ is the peak value of the wave-number distribution in the wave packet; see (78), (79) of \cite{BeGaMaNaTr08_I}. The $\Delta\tau_{\alpha,\beta}$ are the shifts of the reduced arrival times as defined in (99) of \cite{BeGaMaNaTr08_I}. The dynamic and geometric phases acquired by the state with label $\alpha$ from $z=0$ to $z$ are denoted by $\varphi_\alpha(z)$ and $\gamma_\alpha(z)$, respectively. We have
\begin{align}
\varphi_\alpha(z)&=\frac{1}{v_z}\int_0^z\d z'\,\mathcal V_\alpha(z')\ ,\label{phialphanew}\\
\gamma_\alpha(z)&=\I\int_{0}^{z}\d z'\, \lbokp{\alpha(z')}{\frac{\partial}{\partial z'}}{\alpha(z')}\ ,
\end{align}
where $\lbrap{\alpha(z)}$ are the local energy left eigenstates. Note that we use a slightly different notation here, as compared to \cite{BeGaMaNaTr08_I}. To obtain (\ref{Ualpha}) from (101)-(103) of \cite{BeGaMaNaTr08_I} the following replacements have to be made
\begin{align}
\begin{split}
\phi_{\mathrm{dyn},\alpha}&\to\varphi_\alpha(z_a)\ ,\\
\phi_{\mathrm{geom},\alpha}&\to\gamma_\alpha(z_a)\ .
\end{split}
\end{align}

The main quantities of interest to us here are the effective decay rates of the metastable states, see (127) of \cite{GaNaTr2012}, which depend on the path $C$ in parameter space. 
For a state $\alpha\in I$, these decay rates, multiplied by the flight time $T$ from $z_0$ to $z_a$, are given by
\begin{align}\label{Gammaeff}
T\,\Gamma_{\alpha,\mathrm{eff}}(C)=-2\,\mathrm{Im}\,\varphi_\alpha(z_a)+2\,\mathrm{Im}\,\gamma_\alpha(z_a)\ .
\end{align}
The dynamic contribution to $T\,\Gamma_{\alpha,\mathrm{eff}}$ can be written as
\begin{align}
-2\,\mathrm{Im}\,\varphi_\alpha(z_a)&=-\frac{2}{v_z}\int_0^{z_a}\d z\,\mathrm{Im}\,\mathcal V_\alpha(z)\nn\\
&=\frac{m}{\bar k_m}\int_0^{z_a}\d z\,\Gamma_\alpha(z)\label{imphialpha}
\end{align}
and thus depends inversely on $v_z$ and $\bar k_m$, respectively. In (\ref{imphialpha}) $m$ denotes the hydrogen mass. In contrast, the geometric contribution in (\ref{Gammaeff}),
\begin{align}
2\,\mathrm{Im}\,\gamma_\alpha(z_a)\ ,
\end{align}
is independent of $v_z$. This different dependence on $v_z$ allows us to experimentally distinguish between the dynamic and geometric contributions to $T\,\Gamma_{\alpha,\mathrm{eff}}$. For our setup the flight time is
\begin{align}
T=\frac{z_a}{v_z}\approx\frac{0.66}{3500}\,\mathrm{s}\approx 0.2\,\mathrm{ms}\ .
\end{align}
%

%==========================================================================
%==========================================================================
\subsection{Geometric-phase induced lifetime modification}\label{SectionExamples}
\subsubsection{Exemplary field configuration}
\begin{figure}[t]
\centering
\includegraphics[width=0.99\linewidth]{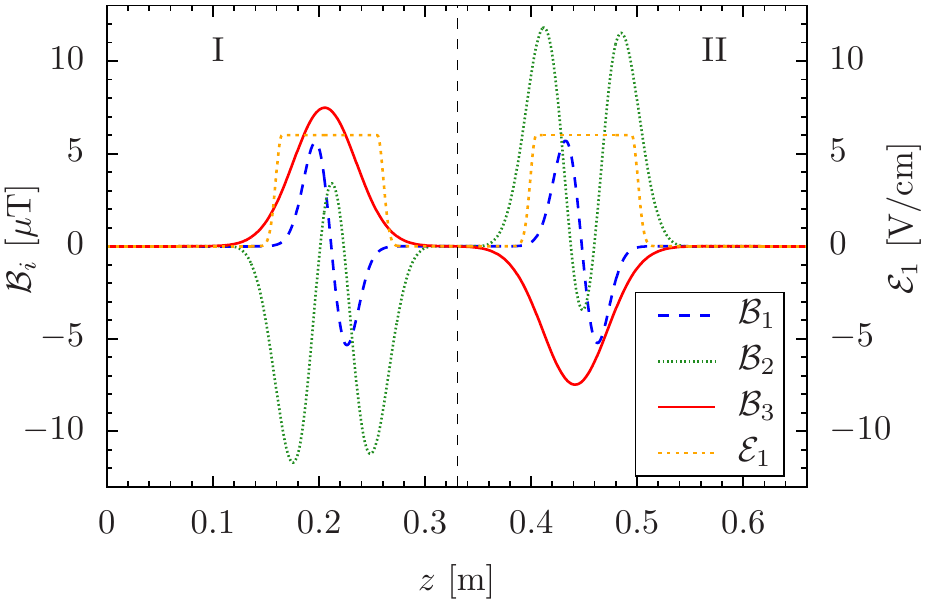}
\caption{The functions $z\mapsto\mathcal E_1(z)$ and $z\mapsto\calBvec(s;z)$ with $s=1$, see Appendix \ref{AppendixField} for details. Ideally, $\calBvec^{\mathrm{rev}}(1,z)=-\calBvec(1;z)$. Regions $\mathrm{I}$ and $\mathrm{II}$ are separated by $z=z_a/2$.}
\label{example1fields}
\end{figure}

\begin{figure}[t]
\centering
\includegraphics[width=0.99\linewidth]{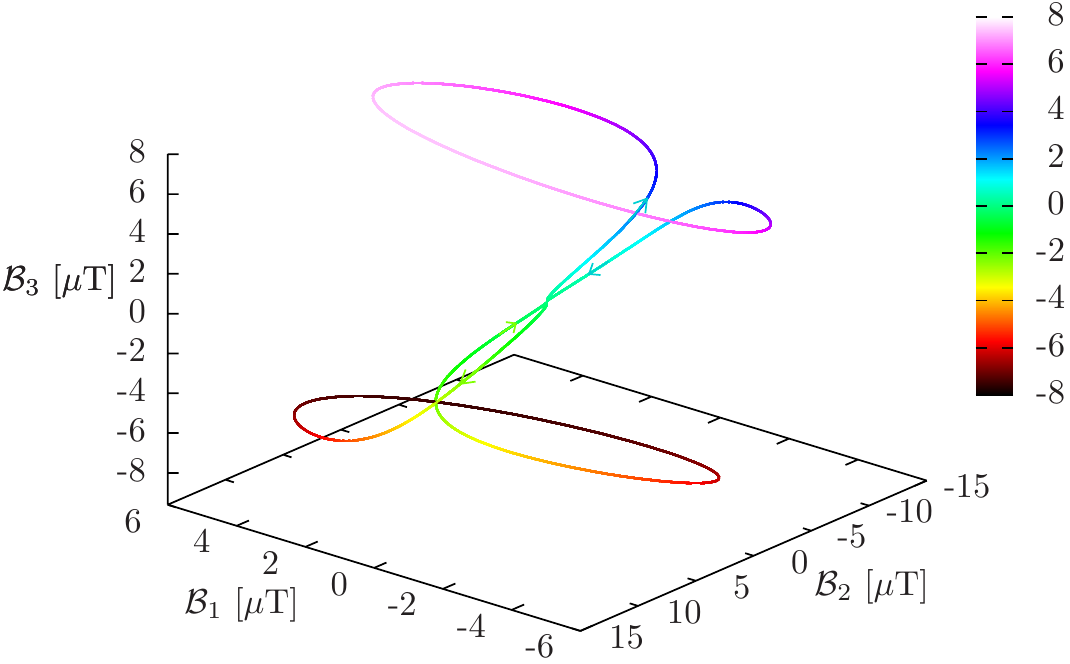}
\ \vspace*{-1.2ex}
\caption{The path $z\mapsto\calBvec(1;z)$ in magnetic field space, starting and ending at $\calBvec=\vec 0$ for $z=0$ and $z_a$, respectively. The values of $\mathcal B_3(1;z)$ are color-encoded. Also $\mathcal E_1(z)$ varies with $z$ as shown in Figure \ref{example1fields} and discussed in the text. The orientation of the path is chosen such that the imaginary parts of the geometric phases are maximised, given the experimental constraints to the magnetic field coils currently available.}
\label{example1Bfields}
\end{figure}

In the following we shall discuss a concrete example of field configurations (\ref{E})-(\ref{zerofields}) and their reverse ones, (\ref{rev}), and calculate the corresponding effective decay rates of metastable H states. We consider the fields shown in Figure \ref{example1fields} (for $s=1$) leading to the path $C$ in parameter space. 
The magnetic part of $C$ is illustrated in Figure \ref{example1Bfields}.
We are looking here for a lifetime shift, that is, a parity conserving (PC), or even effect. We will, therefore, in the following and other than in our previous work \cite{BeGaNa07_I,BeGaNa07_II,BeGaMaNaTr08_I,DeKGaNaTr11,GaNaTr2012}, neglect the very small parity violating (PV) interaction for the hydrogen atom. Hence, in all formulae taken from \cite{BeGaMaNaTr08_I} and \cite{GaNaTr2012}, we leave out the PV contributions.

%=====================================
\captionsetup[figure]{format=plain,font={scriptsize,normalfont}}
\renewcommand{\baselinestretch}{1.0}
\renewcommand{\baselinestretch}{\bls}
\clearcaptionsetup{figure}
%=====================================
As initial and as analysing state we choose the same superposition of the states 9 and 11:
\begin{align}\label{cp}
\begin{split}
c_9=c_{11}&=\frac{1}{\sqrt{2}}\ ,\; c_{10}=c_{12}=0\ ;\\
p_9=p_{11}&=\frac{1}{\sqrt{2}}\ ,\; p_{10}=p_{12}=0\ .
\end{split}
\end{align}

The results shown in the following have been obtained with the help of the numerical software QABSE \cite{DissTB,DissMIT}.
The exemplary path $C$ which we choose in agreement with Eqs.~(\ref{E})-(\ref{zerofields}), represents an external field configuration with electric field components $\mathcal E_1\not=0$, $\mathcal E_2=\mathcal E_3=0$ and magnetic components $\mathcal B_i\not=0$ ($i=1,2,3$). We consider the case where for $s=1$ we have
\begin{align}\label{symmcond}
\begin{split}
\mathcal E_1(z)&=\mathcal E_1(z_a-z)\ ,\\
\calBvec(1;z)&=-\calBvec(1;z_a-z)\ .
\end{split}
\end{align}
That is, we choose $\mathcal E_1(z)$ to be a symmetric function and $\calBvec(1,z)$ to be an antisymmetric function under a reflection at the point $z=z_a/2$.

In Figures \ref{example1fields} and \ref{example1Bfields} we plot the components of these fields as functions of $z$. These fields are inspired  by the realistic design of an actual experimental device, using a fit to calculated and measured field values. The electric field is given in units of V/cm while the magnetic field components are specified in units of $\mu$Tesla. The specific fit functions are listed in Appendix \ref{AppendixField}. We emphasise that these realistic fields satisfy the symmetry conditions (\ref{symmcond}) only to a certain accuracy. We choose the electric field such that $\mathcal E_1(z)=\mathcal E_1(z_a-z)$. The magnetic field is produced by fixed coils, in the regions I and II of the apparatus, one for $\mathcal B_3$ and one for $\mathcal B_1$ and $\mathcal B_2$. The magnetic fields can be varied by changing the currents through these coils. We illustrate the deviations of our field configuration from the ideal symmetric setup (\ref{symmcond}) in Figure \ref{FieldConfigSymmCond}.
In addition to the small violations of (\ref{symmcond}) by the fit functions of Appendix \ref{AppendixField} we have introduced, by hand, a violation of (\ref{symmcond}) by shifting the $z$-component of the magnetic field along the beam axis (dashed line).  As a measure of deviation we use
%=====================================
\captionsetup[figure]{format=plain,font={scriptsize,normalfont}}
\renewcommand{\baselinestretch}{1.0}
\renewcommand{\baselinestretch}{\bls}
\clearcaptionsetup{figure}
%=====================================
%
%
\begin{align}\label{deviation}
\Delta=\frac{1}{z_a}\int_0^{z_a}\d z\,\left\{\sum_{i=1}^3\big[b_i(z)\big]^2\right\}^{1/2}\ ,
\end{align}
where
\begin{align}
b_i(z)=\frac{\mathcal B_i(1;z)+\mathcal B_i(1;z_a-z)}{\underset{0\le z\le z_a}{\mathrm{max}}\mathcal B_i(1;z)}\ .
\end{align}
$\Delta$ vanishes if (\ref{symmcond}) holds. For the field configuration in Figure \ref{example1fields} the deviation (\ref{deviation}) turns out to be $\Delta\approx 8.4\%$ and is mainly due to the asymmetry of $\mathcal B_3$. Note that we deliberately choose the deviations (\ref{deviation}) here almost an order of magnitude larger than in the actual experiment, in order to demonstrate in the following the robustness of our method to this kind of experimental imperfection.

%=====================================
\captionsetup[figure]{format=plain,font={scriptsize,normalfont}}
\renewcommand{\baselinestretch}{1.0}
\begin{figure}[t]
\centering
\includegraphics[width=0.99\linewidth]{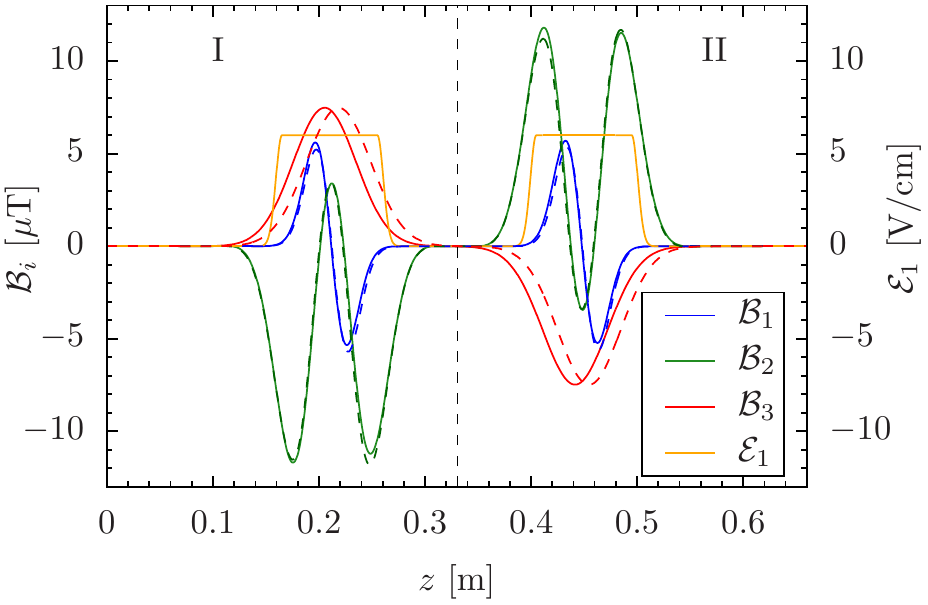}
\caption{Illustration of the deviations of our experimentally motivated field configuration from an ideal configuration satisfying the symmetry conditions (\ref{symmcond}). The solid lines  correspond to the configuration from Figure \ref{example1fields} while the dashed lines indicate the reversed fields, with the sign of the magnetic field switched for presentational purposes, \ie, $-\vec{\mathcal{B}}^{\mathrm{rev}}$.}
\label{FieldConfigSymmCond}
\end{figure}
\renewcommand{\baselinestretch}{\bls}
\clearcaptionsetup{figure}
%=====================================
%
The reverse (\ref{rev}) of the ideal field configuration (\ref{symmcond}), for $s=1$, is obtained by leaving the electric field unchanged and reversing the current through the coils generating the magnetic field,
\begin{align}
\begin{split}\label{rev3}
\mathcal E_1^{\mathrm{rev}}(z)&=\mathcal E_1(z)\ ,\\
\calBvec^{\mathrm{rev}}(1;z)&=-\calBvec(1;z)\ .
\end{split}
\end{align}

While the parameter space in our example is four-dimensional, spanned by $\mathcal E_1$, $\mathcal B_1$, $\mathcal B_2$, $\mathcal B_3$, we can illustrate the projection of the path into the three-dimensional space of the magnetic fields. 
Figure \ref{example1Bfields} shows this projection of the path $C_s$ (\ref{path}) for $s=1$. The corresponding $z$-dependence of $\mathcal E_1(z)$ is as shown in Figure \ref{example1fields}. That is, $\mathcal E_1(z)$ starts at zero and is positive when $\calBvec(z)$ traces out the upper loop in Figure \ref{example1Bfields}. After this, $\mathcal E_1(z)$ goes to zero at $z=z_a/2$ before becoming positive again while $\calBvec(z)$ traces out the lower loop in Figure \ref{example1Bfields}. Finally, both $\mathcal E_1(z)$ and $\calBvec(z)$ go back to zero before ending at $z=z_a$.

\begin{figure}[tbp]
\centering
\includegraphics[width=0.99\linewidth]{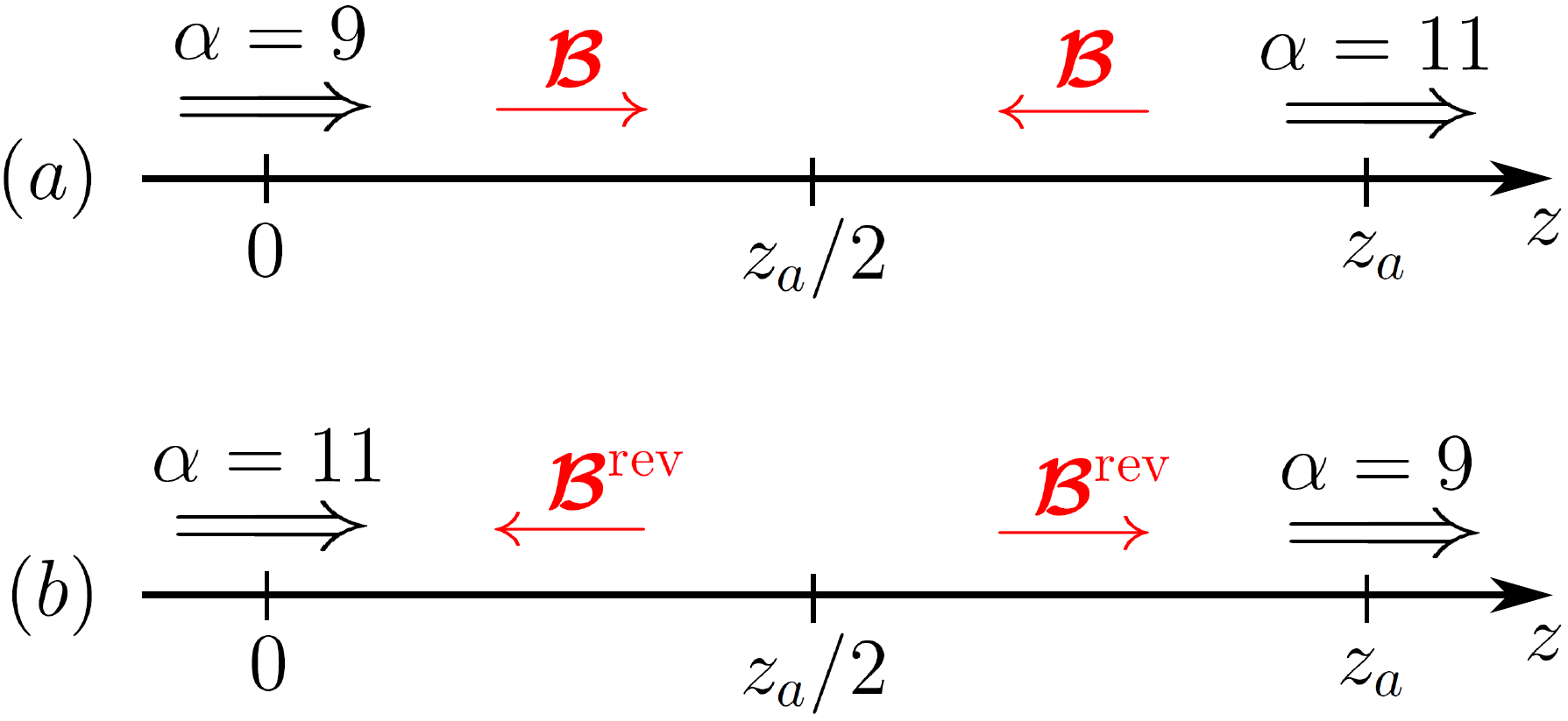}
\caption{Illustration of the renumbering of states in the case that only $\mathcal B_3(s;z)\not=0$. The double line arrows indicate the spin directions. In (a) the state starting at $z=0$ with label $\alpha=9$  is subject to the path $C_s$ in parameter space and arrives with label $\alpha=11$. In the reverse field configuration (b) the corresponding state to start with has label $\alpha=11$ and is relabeled as $\alpha=9$ for $z>z_{a}/2$.}
\label{spinflip}
\end{figure}

The evolution of the states in the interferometer should be adiabatic wherever geometric phases are picked up for $0<z<z_a/2$ and $z_a/2<z<z_a$. We have made sure that this is true for all cases considered; see Appendix \ref{AppendixAdiabat}. The point $z=z_a/2$ is special since there we have $\calEvec=0$ and $\calBvec=0$ as required in (\ref{zerofields}), implying a degeneracy to appear at this point. Making use of the numbering scheme as explained in Appendix A of \cite{GaNaTr2012} we find that a state with label $\alpha=9$ ($\alpha=11$) entering from $z<z_a/2$ will have the label $\alpha=11$ ($\alpha=9$) for $z>z_a/2$. Hereby, we make sure that the phases of the states are continuous for $z=z_a/2$ despite their renumbering. In the following we shall, therefore, label the states, energies, etc., with $\9$ and $\1$ where the first/second number corresponds to the label $\alpha$ in the first/second half of the interferometer. Note that for the states $\alpha=10$ and $12$ there is no relabelling at $z=z_a/2$. Note furthermore that, when switching from the path defined by the fields (\ref{E}), (\ref{B}) to the reverse path (\ref{rev}), we have to compare the states $\9$ with $\1$ and, correspondingly, $\1$ with $\9$. This becomes particularly clear if in (\ref{E}), (\ref{B}) we consider a path with only $\mathcal B_3(s;z)\not=0$, of the form shown in Figure \ref{example1fields} and with $\mathcal B^{\mathrm{rev}}_3(s;z)=\mathcal B_3(s;z_a-z)$. The states $\alpha=9$ ($\alpha=11$) are then those with spin parallel (antiparallel) to $\calBvec$. The renumbering is  illustrated in Figure \ref{spinflip}, for the system in state $\alpha=\9$ within a field configuration path $C_s$ and in the corresponding state $\alpha=\1$ within $\overline C_s$. 

%=====================================
\captionsetup[figure]{format=plain,font={scriptsize,normalfont}}
\renewcommand{\baselinestretch}{1.0}
\renewcommand{\baselinestretch}{\bls}
\clearcaptionsetup{figure}
%=====================================

\subsubsection{Dynamic and geometric phases}
\begin{figure}[t]
\centering
\includegraphics[width=0.99\linewidth]{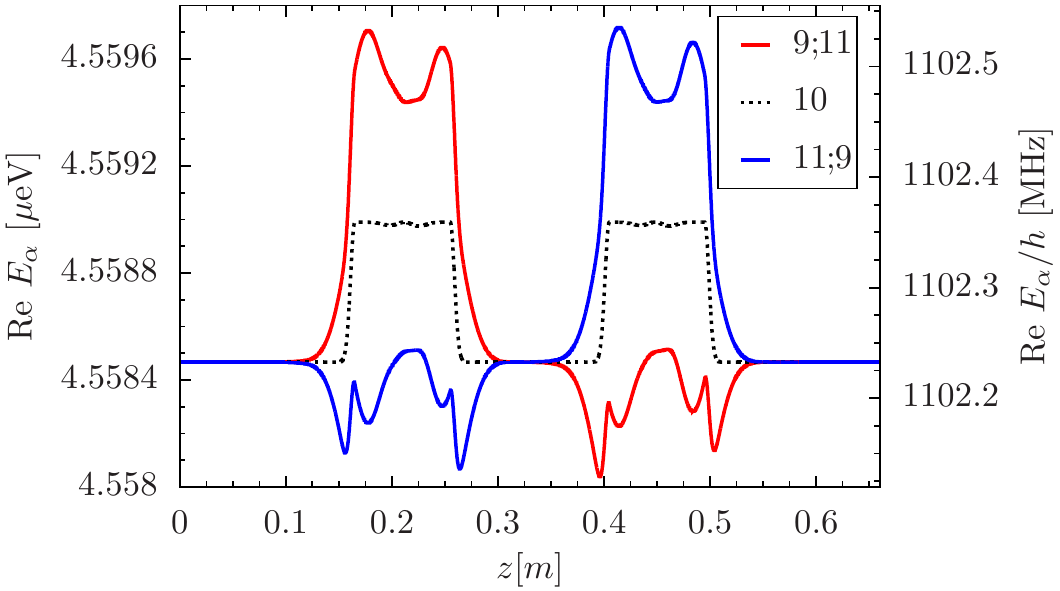}
\caption{The real parts of the energies $E_\alpha(z)$ of the atomic states $\alpha=\9$, $10$, and $\1$ in the fields shown in Figure \ref{example1fields}, with $s=1$.}
\label{example1energies}
\end{figure}

The dynamical phases picked up by the states traversing the external field configurations are defined by the $z$-dependencies of their eigenenergies. 
In Figure \ref{example1energies} we show, for $s=1$, the real parts of the energies $E_\alpha(z)$ for $\alpha=\9,10,\1$, exhibiting the Zeeman- and Stark-shifts according to the fields shown in Figure \ref{example1fields}. As we can see from (73) of \cite{GaNaTr2012} the functional dependence of $E_\alpha(z)$ on the external fields is as follows:
\begin{align}\label{Ealpha}
E_\alpha(z)\equiv E_\alpha\big(\calEvec^2(z),\calBvec^2(z),[\calEvec(z)\cdot\calBvec(z)]^2\big)\ .
\end{align}
For our field configurations this can be simplified to
\begin{align}\label{ourfield}
E_\alpha(z)=E_\alpha\big([\mathcal E_1(z)]^2,\calBvec^2(z),[\mathcal E_1(z)\mathcal B_1(z)]^2\big)\ .
\end{align}
We find, therefore, that in the ideal case where (\ref{rev3}) holds the eigenenergies are the same, taking $s=1$, for the field path $C_{1}$ and the reverse path $\overline C_1$,
\begin{align}
\left.E_\alpha(z)\right|_{C_1}=\left.E_\alpha(z)\right|_{\overline{C}_1}\ .
\end{align}
The same holds for the effective potential $\mathcal V_\alpha(z)$ because the additional geometric contributions are negligible, see (\ref{ReVE}) and (\ref{ImVE}),
\begin{align}\label{effpotCCbar}
\left.\mathcal V_\alpha(z)\right|_{C_1}=\left.\mathcal V_\alpha(z)\right|_{\overline{C}_1}\ .
\end{align}
For the dynamic phases $\varphi_\alpha(z)$ we have, therefore, from (\ref{phialphanew}) and (\ref{effpotCCbar}) again in the ideal case
\begin{align}\label{phiCCbar}
\left.\varphi_\alpha(z)\right|_{C_1}=\left.\varphi_\alpha(z)\right|_{\overline{C}_1}\ .
\end{align}
In (\ref{Ealpha})--(\ref{phiCCbar}) we have
\begin{align}
\alpha\in\{\9,\1,10,12\}\ .
\end{align}
%
%
%=====================================
\captionsetup[figure]{format=plain,font={scriptsize,normalfont}}
\renewcommand{\baselinestretch}{1.0}
\renewcommand{\baselinestretch}{\bls}
\clearcaptionsetup{figure}
%=====================================
%

In Figure \ref{example1energies} we show $\mathrm{Re}\,E_\alpha(z)$ for the realistic field configuration of Figure \ref{example1fields} where the symmetry relations (\ref{symmcond}) hold only approximately. In case that (\ref{symmcond}) would hold exactly the red curve ($\alpha=\9$) would be the reflection of the blue curve ($\alpha=\1$) on $z=z_a/2$. We see that this reflection symmetry holds to a good approximation. The observed asymmetry in Figure \ref{example1energies} is caused mainly by the shift of $-\mathcal B_3^{\mathrm{rev}}$ {\wrt} $\mathcal B_3$, see Figure \ref{FieldConfigSymmCond}, but does not qualitatively affect the main findings of this work. The asymmetry should rather be regarded as a realistic complication which our methods can easily deal with.
The difference
%
%
%=====================================
\captionsetup[figure]{format=plain,font={scriptsize,normalfont}}
\renewcommand{\baselinestretch}{1.0}
\begin{figure}[t]
\centering
\includegraphics[width=0.99\linewidth]{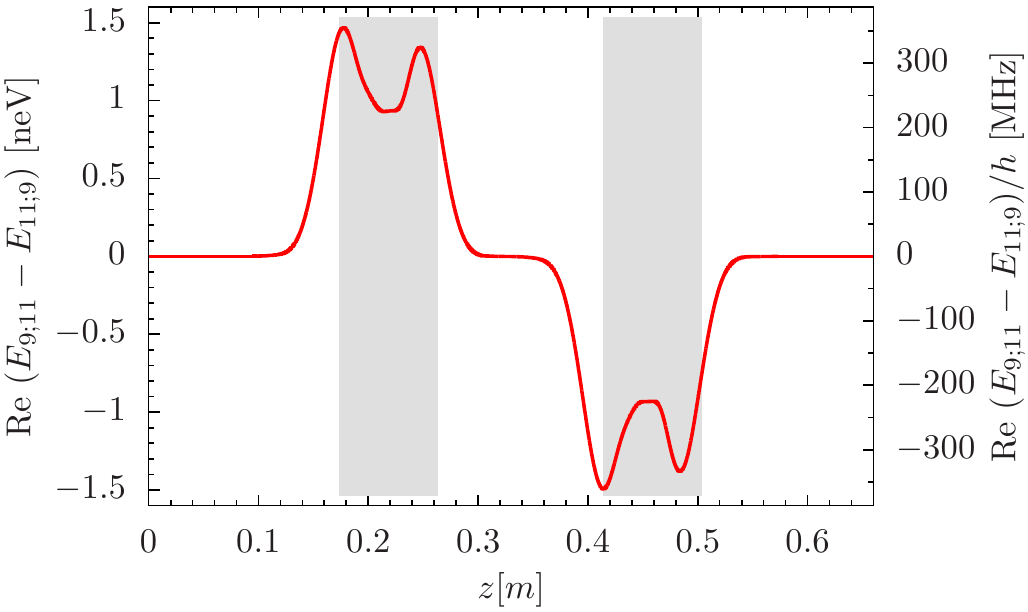}
\caption{The real part of the energy difference (\ref{Ediff}) for the fields shown in Figure \ref{example1fields}, with $s=1$. The shaded areas indicate the regions of non-zero electric fields.}
\label{example1energydifferences}
\end{figure}
\renewcommand{\baselinestretch}{\bls}
\clearcaptionsetup{figure}
%=====================================
%=====================================
\captionsetup[figure]{format=plain,font={scriptsize,normalfont}}
\renewcommand{\baselinestretch}{1.0}
\begin{figure}[t]
\centering
\includegraphics[width=0.8\linewidth]{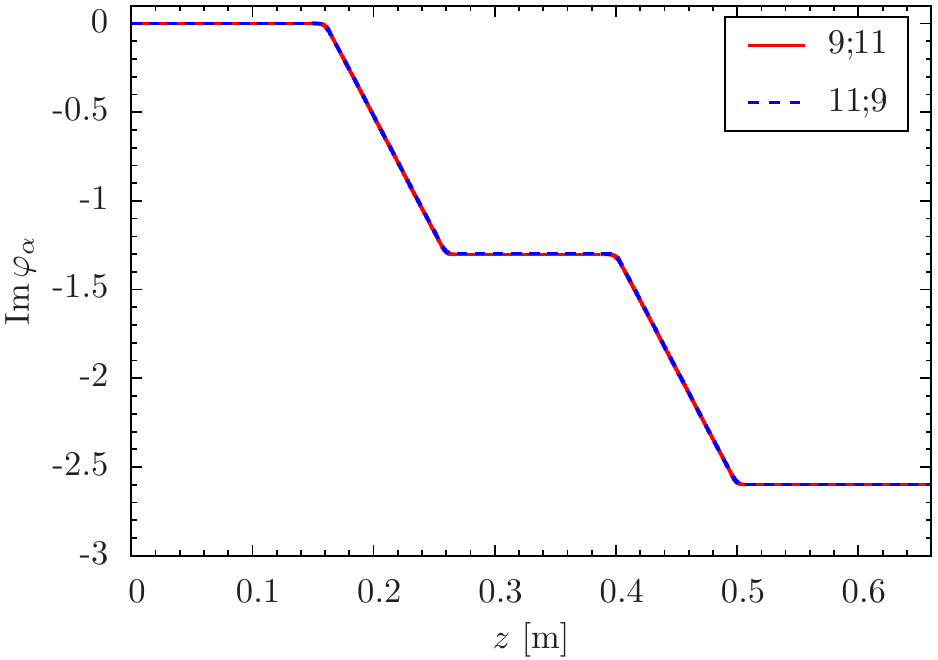}
\hspace*{4.5ex}\
\caption{The imaginary parts of the dynamic phase (\ref{imphialpha}), $\mathrm{Im}\,\varphi_\alpha(z)$, as function of $z$, for $s=1$, for the states $\alpha=\9$ and $\1$. The field configuration is given in Figure \ref{example1fields}. The plot clearly shows where the imaginary parts of the dynamic phases are picked up along the $z$-axis.}
\label{example1imag}
\end{figure}
\renewcommand{\baselinestretch}{\bls}
\clearcaptionsetup{figure}
%=====================================
%
\begin{align}\label{Ediff}
\mathrm{Re}\,\big[E_{\9}(z)-E_{\1}(z)\big]\ ,
\end{align}
again for $s=1$, is shown in Figure \ref{example1energydifferences}. The adiabaticity conditions associated with these energy differences can be checked easily; see Appendix \ref{AppendixAdiabat}. 

In Figure \ref{example1imag} we show the $z$-dependent imaginary parts of the dynamic phase for the states $\alpha=\9$ and $\1$ exposed to the fields in Figure \ref{example1fields} where $s=1$. For these fields the imaginary parts of the dynamic phases are, within the accuracy of our numerical calculations, the same for $\alpha=\9$ and $\alpha=\1$. For the reverse field configuration (\ref{rev}), again with $s=1$ and in the ideal case where (\ref{rev3}) holds, the imaginary parts of the dynamic phases are the same as for the original field configuration; see (\ref{phiCCbar}).

%=====================================
\captionsetup[figure]{format=plain,font={scriptsize,normalfont}}
\renewcommand{\baselinestretch}{1.0}
\begin{figure}[t]
\centering
\includegraphics[width=0.82\linewidth]{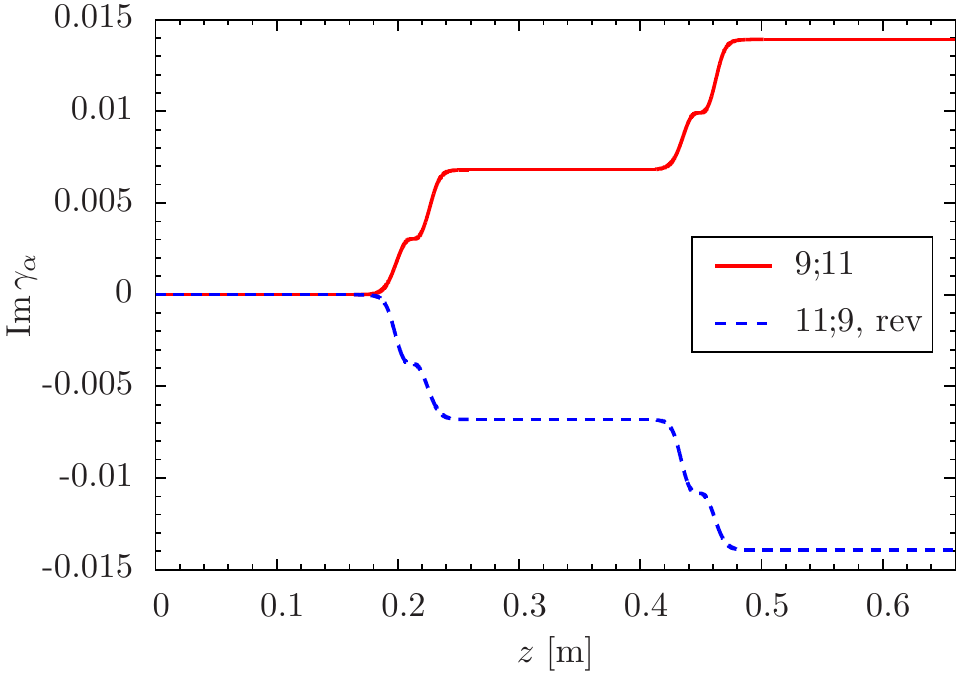}
\hspace*{4.5ex}\
\caption{The imaginary part $\mathrm{Im}\,\gamma_\alpha(z)$ of the geometric phase, as function of $z$, for $s=1$, for the state $\alpha=\9$ in the field configuration path $C_1$ given in Figure \ref{example1fields}, and for the state $\alpha=\1$ within the reversed configuration $\overline{C}_{1}$. The curves are identical for the states $\alpha=\1$, with $C_{1}$, and for $\9$, with $\overline{C}_{1}$.}
\label{example1imagdiff}
\end{figure}
\renewcommand{\baselinestretch}{\bls}
\clearcaptionsetup{figure}
%=====================================

In Figure \ref{example1imagdiff} we show the imaginary parts of the geometric phases, $\mathrm{Im}\,\gamma_{\alpha}(z)$, as functions of $z$ for $\alpha=\9$ and the curve $C_{1}$, and for $\alpha=\1$ and $\overline{C}_{1}$.
A clear difference in $\mathrm{Im}\,\gamma_{\alpha}(z)$ between these two cases can be seen.
For the curve $C_{1}$ the results for $\alpha=\9$ and $\1$ are the same.
This is also the case for the curve $\overline{C}_{1}$.
Note that here and in the following we compare $\alpha=\9$  ($\1$) in the field path $C_{s}$ to $\alpha=\1$  ($\9$) in the field path $\overline{C}_{s}$, thus taking into account the label change explained in Figure \ref{spinflip}.
The sign change of $\mathrm{Im}\,\gamma_{\alpha}(z_{a})$ when going from $\alpha=\9$ and $C_{1}$ to $\alpha=\1$ and $\overline{C}_{1}$ in Figure \ref{example1imagdiff} is clear from the property of geometric phases as line integrals.
The fact that we have the same result for $\mathrm{Im}\,\gamma_{\alpha}(z_{a})$ for $\alpha=\9$ and $\1$ is due to the special configuration of fields chosen; see Figures \ref{example1fields} and \ref{example1Bfields}.

\begin{figure}[t]
\centering
\includegraphics[width=0.92\linewidth]{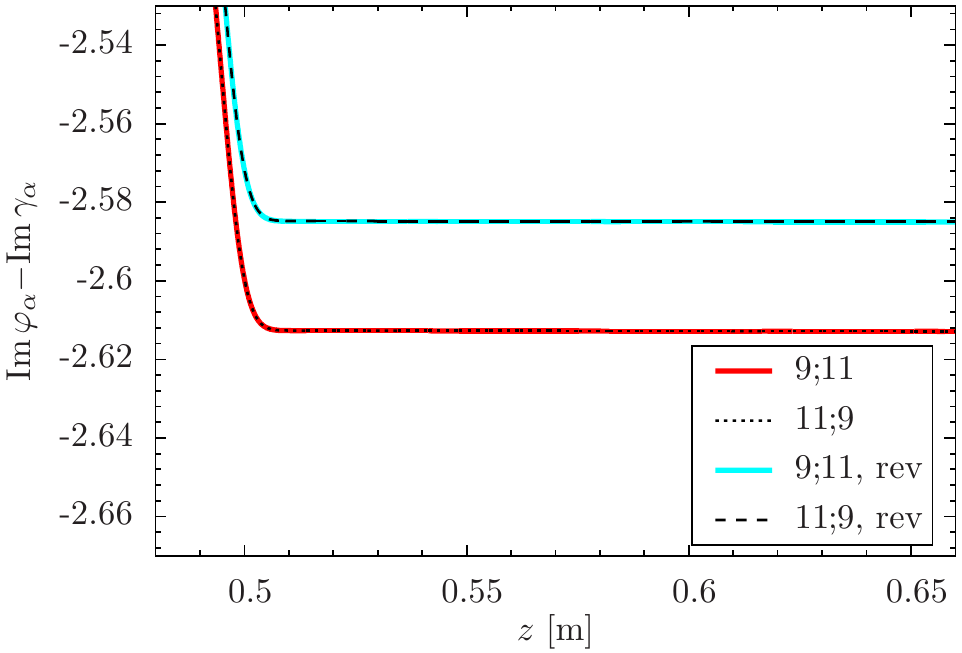}
%\hspace*{4.5ex}\
\caption{Combining the data shown in Figures \ref{example1imag} and \ref{example1imagdiff}, we depict $\mathrm{Im}\,\varphi_\alpha(z)-\mathrm{Im}\,\gamma_\alpha(z)$ for both the states $\alpha=\9$ and $\alpha=\1$ in $C_{1}$, together with the data obtained with the reversed path $\overline{C}_1$. The differences $\mathrm{Im}\,\varphi_\alpha(z_a)-\mathrm{Im}\,\gamma_\alpha(z_a)$ at the end of the interferometer are used in (\ref{example1Gammaeff}) to extract the lifetime modification (\ref{R}), (\ref{Ra}).}
\label{figInset}
\end{figure}

We now turn to the difference of the imaginary parts of the dynamic and geometric phases. For the field configuration of Figure \ref{example1fields}, corresponding to $s=1$ and the path $C_1$ in parameter space, this difference is shown in Figure \ref{figInset} for $\alpha=\9$ and $\alpha=\1$. For the path $C_1$ and the reverse path $\overline{C}_1$ we see a clear difference in $\mathrm{Im}\,\varphi_\alpha(z)-\mathrm{Im}\,\gamma_\alpha(z)$. For the effective decay rates multiplied by the flight times, see (\ref{Gammaeff}), we get
\begin{align}\label{example1Gammaeff}
%\begin{split}
T\,\Gamma_{\9,\mathrm{eff}}(C_1)&=T\,\Gamma_{\1,\mathrm{eff}}(C_1)
\nonumber\\
&=\left.\left(-2\mathrm{Im}\,\varphi_{\9}(z_a)+2\mathrm{Im}\,\gamma_{\9}(z_a)\right)\right|_{C_1}
\nonumber\\
&=2(2.599+0.0139)\ ,
\nonumber\\[0.6ex]
T\,\Gamma_{\9,\mathrm{eff}}(\overline{C}_1)&=T\,\Gamma_{\1,\mathrm{eff}}(\overline{C}_1)
\nonumber\\
&=\left.\left(-2\mathrm{Im}\,\varphi_{\9}(z_a)+2\mathrm{Im}\,\gamma_{\9}(z_a)\right)\right|_{\overline{C}_1}
\nonumber\\
&=2(2.599-0.0139)\ ,
%\end{split}
\end{align}
if the symmetry condition (\ref{symmcond}) is satisfied. The latter implies that a maximum revival, that is, a spin echo, can be observed at $s=1$, and the maxima of $\mathcal F_p(C_s)$ and $\mathcal F_p(\overline C_s)$ are both found at $s=1$. Furthermore, the same decay rates (\ref{example1Gammaeff}) are obtained for atomic states initially prepared in any superposition of $\alpha=9$ and $\alpha=11$. From the values (\ref{example1Gammaeff}) we obtain the ratio $R_\alpha$ of the fluxes of metastable hydrogen atoms in states $\alpha=\1$ and field-path $\overline{C}_{1}$ and $\alpha=\9$ and path ${C}_1$ as
\begin{align}\label{R}
R_{\9}=\frac{\exp[-T\,\Gamma_{\1,\mathrm{eff}}(\overline{C}_1)]}{\exp[-T\,\Gamma_{\9,\mathrm{eff}}(C_1)]}=1.057\ .
\end{align}
Similarly we get
\begin{align}\label{Ra}
R_{\1}=\frac{\exp[-T\,\Gamma_{\9,\mathrm{eff}}(\overline{C}_1)]}{\exp[-T\,\Gamma_{\1,\mathrm{eff}}(C_1)]}=1.057\ .
\end{align}
%
%=====================================
\captionsetup[figure]{format=plain,font={scriptsize,normalfont}}
\renewcommand{\baselinestretch}{1.0}
\renewcommand{\baselinestretch}{\bls}
\clearcaptionsetup{figure}
%=====================================

%=====================================
\captionsetup[figure]{format=plain,font={scriptsize,normalfont}}
\renewcommand{\baselinestretch}{1.0}
\begin{figure}[t]
\centering
\includegraphics[width=0.92\linewidth]{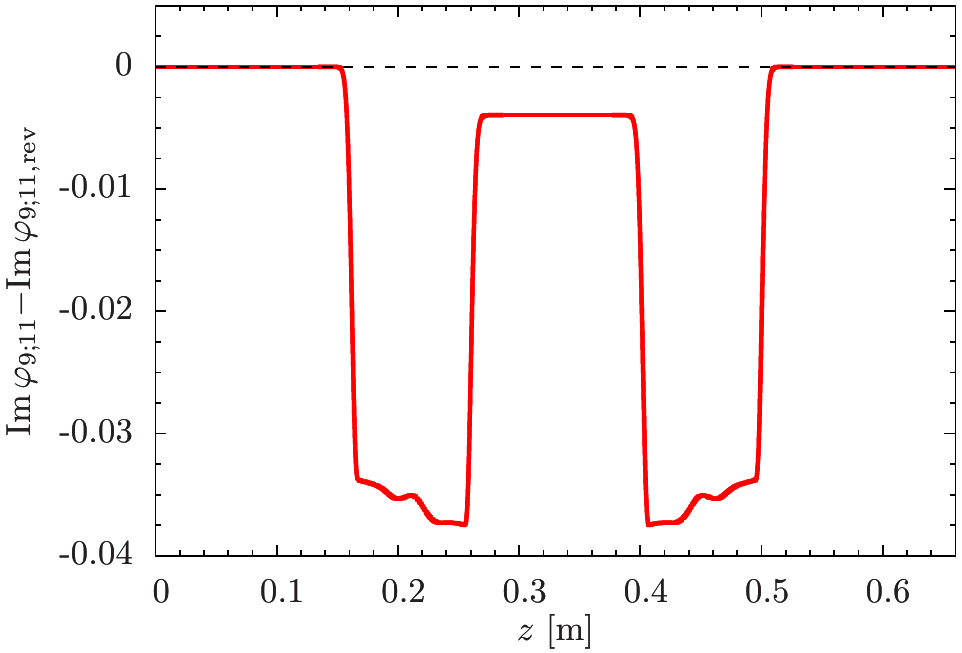}
%\vspace*{2.0ex}\\
\caption{The difference $\mathrm{Im}\,\varphi_{\9}(z)-\mathrm{Im}\,\varphi_{\9,\mathrm{rev}}(z)$ as a function of $z$, using the field configuration $C_1$ in Figure \ref{example1fields} and its reverse $\overline C_1$. This difference vanishes for fields obeying the symmetry condition (\ref{symmcond}) and is a measure for the violation of (\ref{phiCCbar}). As for spin echo signals, however, the $z$-dependence only enters at $z=z_a$, the violation of (\ref{phiCCbar}) is of no concern here. This makes our method rather robust with respect to imperfections in the experimental field configurations of the type (\ref{deviation}).}
\label{figimDynDiff9}
\end{figure}
\renewcommand{\baselinestretch}{\bls}
\clearcaptionsetup{figure}
%=====================================
We expect the effect on the atomic lifetimes which is at the level of more than 5\% to be accessible in a realistic experiment. However, $R_\alpha$ in (\ref{R}), (\ref{Ra}) is an appropriate measure of geometric lifetime modification only if a \textit{symmetric} field configuration according to (\ref{symmcond}) is given. As we shall see below, the maxima of $\mathcal F_p(C_s)$ and $\mathcal F_p(\overline C_s)$ are in general found at different values of $s$ if (\ref{symmcond}) is not satisfied exactly. Although the norm of the atomic states $\alpha=\9$ and $\alpha=\1$ decays as obtained from (\ref{example1Gammaeff}), an initial superposition of $\alpha=\9$ and $\alpha=\1$ travelling through an asymmetric field configuration leads to interference patterns for which the maximal revival of the initial state is not reached at $s=1$. If the deviation from (\ref{symmcond}) were large enough, even completely destructive interference could be observed, misleadingly indicating large decay rates. Therefore, we cannot extract the lifetime modification for our slightly asymmetric realistic fields by only comparing $\mathcal F_p(C_1)$ and $\mathcal F_p(\overline C_1)$. 
Deviations from the symmetry conditions (\ref{symmcond}) occurring in realistic situations, however, do not affect the spin-echo measurements we are proposing here.
To demonstrate this we show in Figure \ref{figimDynDiff9} the difference of the imaginary parts of $\varphi_{\9}(z)$ and $\varphi_{\9,\mathrm{rev}}(z)$ for $s=1$ where the reversed fields are the realistic ones fulfilling (\ref{symmcond}) only approximately; see Figure \ref{FieldConfigSymmCond}. We see that  $\mathrm{Im}\,\varphi_{\9}(z)-\mathrm{Im}\,\varphi_{\9,\mathrm{rev}}(z)$ \textit{is} different from zero, but for $z=z_a$ the difference vanishes, since the integral over both regions I and II in (\ref{deviation}) is the same. For our lifetime measurements only the value of these imaginary parts at $z=z_a$ matters and, therefore, our results (\ref{example1Gammaeff}), (\ref{R}) and (\ref{Ra}) hold unchanged also for our realistic case where (\ref{symmcond}) is satisfied only approximately.

\subsubsection{Spin-echo measurement procedure}
We now turn to the actual measurement to be done with the spin-echo apparatus in order to extract the lifetime differences calculated above. A direct measurement of (\ref{R}), (\ref{Ra}) with the spin-echo field configuration in Figure \ref{example1fields} is possible by starting with hydrogen in the state $\alpha=9$ and projecting onto $\alpha=11$, \ie, $c_9=p_{11}=1$. 
The results obtained should then be compared to the case with reversed fields, starting with state $\alpha=11$ and projecting onto $\alpha=9$ at $z=z_{a}$.
Notice, hereby, the change of labeling of the states at $z=z_a/2$; see Figure \ref{spinflip} and the discussion after (\ref{rev3}). However, aiming at an actual spin-echo measurement, we propose to choose identical initial and analysing states, \ie, the superpositions in (\ref{cp}).

\begin{figure}[t]
\centering
\includegraphics[width=0.86\linewidth]{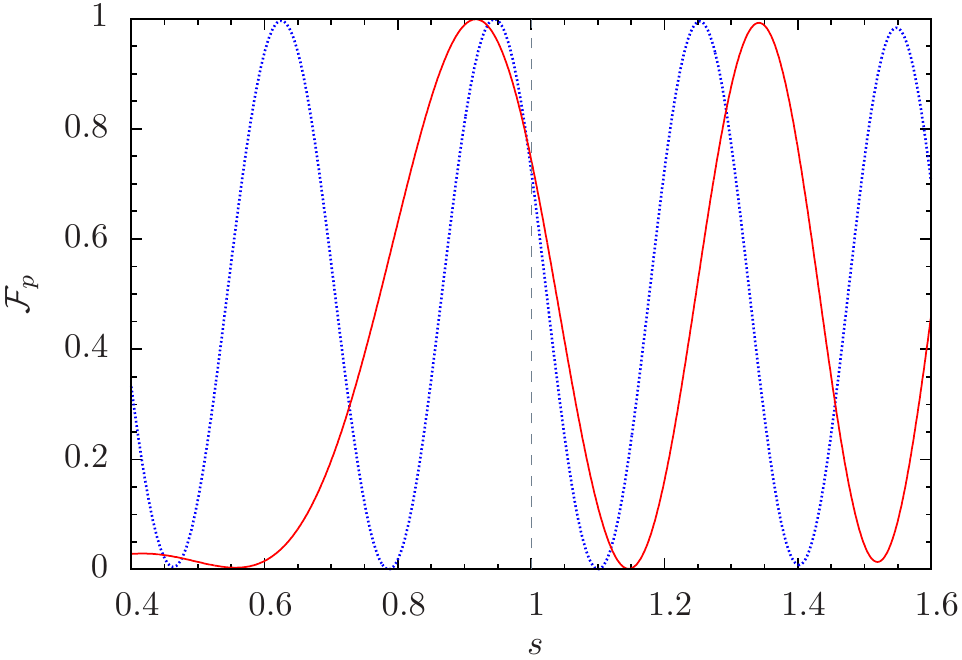}
\vspace*{-0.3ex}
\caption{Spin echo integrated-flux curves for the paths $C_s$ (red solid line) and $\overline C_s$ (blue dotted line) using (\ref{cp}) and the field configuration in Figure \ref{example1fields}, but with the electric field set to zero. Experimentally, $s$ can be varied by varying the current through the coil which generates the $\mathcal B_3$-field in the second (first) half of the interferometer for $C_s$ ($\overline C_s$). The vertical dashed line marks $s=1$. Without electric field the decay of the metastable states is negligible, and the spin echos reach almost unit amplitude for several values of $s$. However, the amplitudes also depend on the real parts of the $s$-dependent dynamic and geometric phases, as does the separation of the maxima along the $s$-axis.}
\label{example1zeroE1SpinechoANDrev}
\end{figure}

\begin{figure}[t]
\centering
\includegraphics[width=0.81\linewidth]{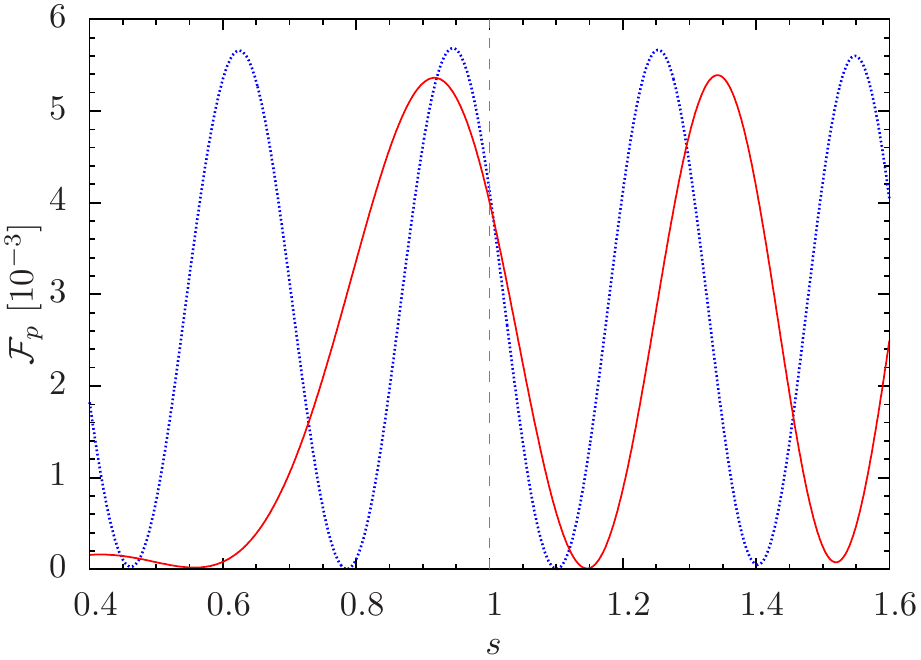}
\caption{Spin echo curves for the paths $C_s$ (red solid line) and $\overline C_s$ (blue dotted line) using (\ref{cp}) as in Figure \ref{example1zeroE1SpinechoANDrev}, but with the electric field turned on. The vertical dashed line marks $s=1$. The electric field results in decreased amplitudes of the spin echo curves, but the general shapes of the interference patterns are unchanged, cf.~Figure~\ref{example1zeroE1SpinechoANDrev}. However, the presence of the electric field allows for the imaginary geometric phase to emerge after the closed path shown in Figure \ref{example1Bfields} has been traced out in parameter space, resulting in different values of the heights of maxima when comparing $C_s$ with $\overline C_s$. See Figure \ref{example1SpinechoANDrevs1} for an enlarged display of the region around $s=1$.}
\label{example1SpinechoANDrev}
\end{figure}

Varying $s$, we obtain the spin-echo curves shown in Figures \ref{example1zeroE1SpinechoANDrev} and \ref{example1SpinechoANDrev}. These plots conveniently illustrate how lifetime modifications through geometric phases can be observed experimentally. The magnitude of this effect can be easily extracted by comparing the amplitudes of the spin-echo curves measured for $C_s$ and $\overline C_s$ as we discuss in more detail in the following.

%=====================================
\captionsetup[figure]{format=plain,font={scriptsize,normalfont}}
\renewcommand{\baselinestretch}{1.0}
\renewcommand{\baselinestretch}{\bls}
\clearcaptionsetup{figure}
%=====================================
%=====================================
\captionsetup[figure]{format=plain,font={scriptsize,normalfont}}
\renewcommand{\baselinestretch}{1.0}
\renewcommand{\baselinestretch}{\bls}
\clearcaptionsetup{figure}
%=====================================

\begin{figure}[t]
\centering
\includegraphics[width=0.9\linewidth,height=0.65\linewidth]{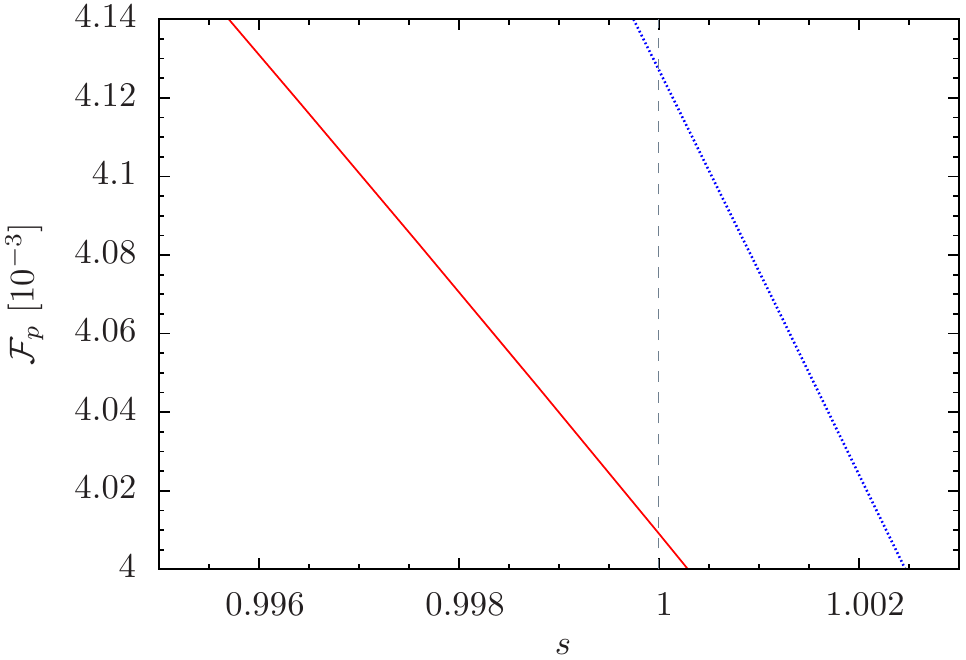}
\caption{Magnification of the spin echo integrated-flux curves shown in Fig.~\ref{example1SpinechoANDrev}, for the paths $C_s$ (red solid line) and $\overline C_s$ (blue dotted line) using (\ref{cp}) near $s=1$. The differing lifetimes specified in (\ref{example1Gammaeff}) show up as different values of the spin echo fluxes for $C_1$ and $\overline C_1$, respectively. However, while the reversed path gives a lower decay rate than the red path at $s=1$, the spin echo signal is not entirely determined by its amplitude, and we have to take into account the frequency of oscillation due to the cosine in (\ref{Fpapprox}).}
\label{example1SpinechoANDrevs1}
\end{figure}

Figure \ref{example1SpinechoANDrevs1} shows the behaviour of $\mathcal F_p(C_s)$ and $\mathcal F_p(\overline C_s)$ near $s=1$ in an enlarged scale. The lifetime differences due to the differing imaginary parts of the geometric phases for $C_1$ and $\overline C_1$ cause different spin echo curves for $C_1$ and $\overline C_1$. 
Note, however, that for a quantitative analysis we have to take into account also the real parts of the dynamic and geometric phases as will be explained below.

%=====================================
\captionsetup[figure]{format=plain,font={scriptsize,normalfont}}
\renewcommand{\baselinestretch}{1.0}
\renewcommand{\baselinestretch}{\bls}
\clearcaptionsetup{figure}
%=====================================
As our main result we predict that the amplitudes of the spin-echo signals obtained for $C_s$ and $\overline C_s$ differ due to imaginary geometric phases, to an extent that the effect is large enough to be experimentally accessible. The effect is extracted from the main features of the interference patterns $\mathcal F_p(C_s)$ and $\mathcal F_p(\overline C_s)$, with and without the electric field component $\mathcal E_1$ as shown in Figure \ref{example1fields}. Comparing Figures \ref{example1zeroE1SpinechoANDrev} and \ref{example1SpinechoANDrev}, we observe a decreased amplitude as the most pronounced effect of the electric field, while the phase of the interference patterns is not visibly affected, \ie, the electric field has negligible influence on the real parts of the geometric phases. 

The frequencies of $\mathcal F_p(C_s)$ and $\mathcal F_p(\overline C_s)$ in Figure \ref{example1zeroE1SpinechoANDrev} \wrt $s$ are distinctly different, and both are $s$-dependent. As we will discuss in the following, the behavior of $\mathcal F_p$ as a function of $s$ is easily understood in terms of the $s$-dependent phases since the field configuration in Figure \ref{example1fields} allows for simplifications of the general expression (\ref{Fp}). It will become clear that the different $s$-dependences of $\mathcal F_p(C_s)$ and $\mathcal F_p(\overline C_s)$ result from an interference effect involving the real parts of the geometric phases, while the different values of the maxima of $\mathcal F_p(C_s)$ and $\mathcal F_p(\overline C_s)$ originate mainly from the differences in the imaginary parts of the geometric phases.

\begin{figure}[t]
\centering
\includegraphics[width=0.9\linewidth]{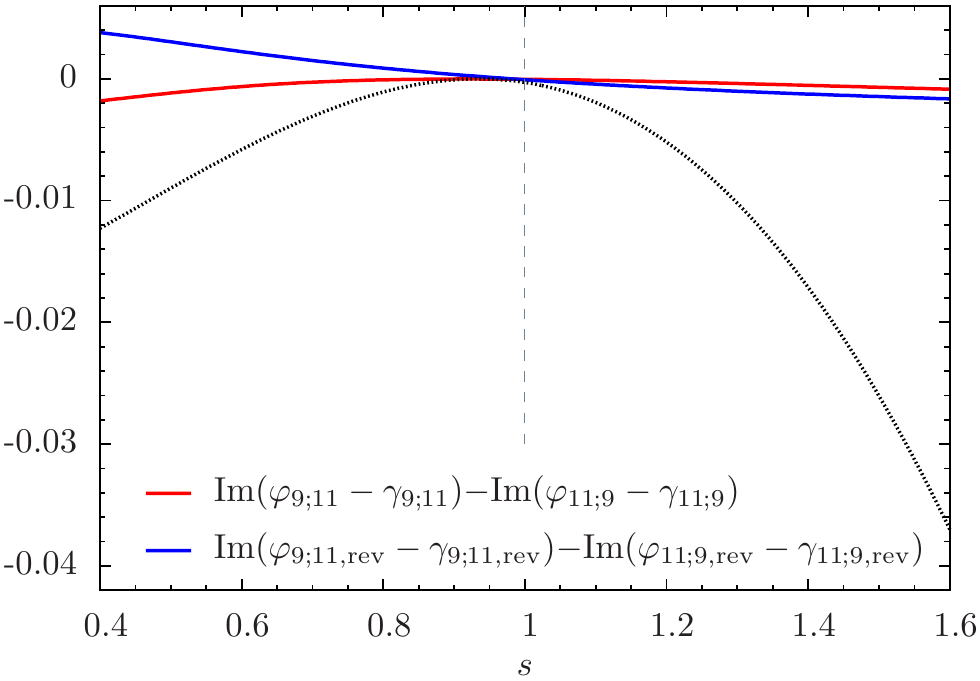}
\ \vspace*{1ex}
\caption{The relevant quantities that allow for the approximation (\ref{Fpapprox}) of (\ref{Fp}), given as functions of $s$ for the field configuration in Figure \ref{example1fields} and its reverse. The black dotted line shows $\exp [-(\Delta\tau_\beta-\Delta\tau_\alpha)^2/(8\sigma'^2_k)]-1$ for $C_s$ and $\beta=\9$ and $\alpha=\1$. The curves for $\beta=\1$ and $\alpha=\9$ as well as for the reversed path $\overline C_s$ are the same.}
\label{SpinEchoPuzzle20141121data2}
\end{figure}

\begin{figure}[t]
\centering
\includegraphics[width=0.865\linewidth]{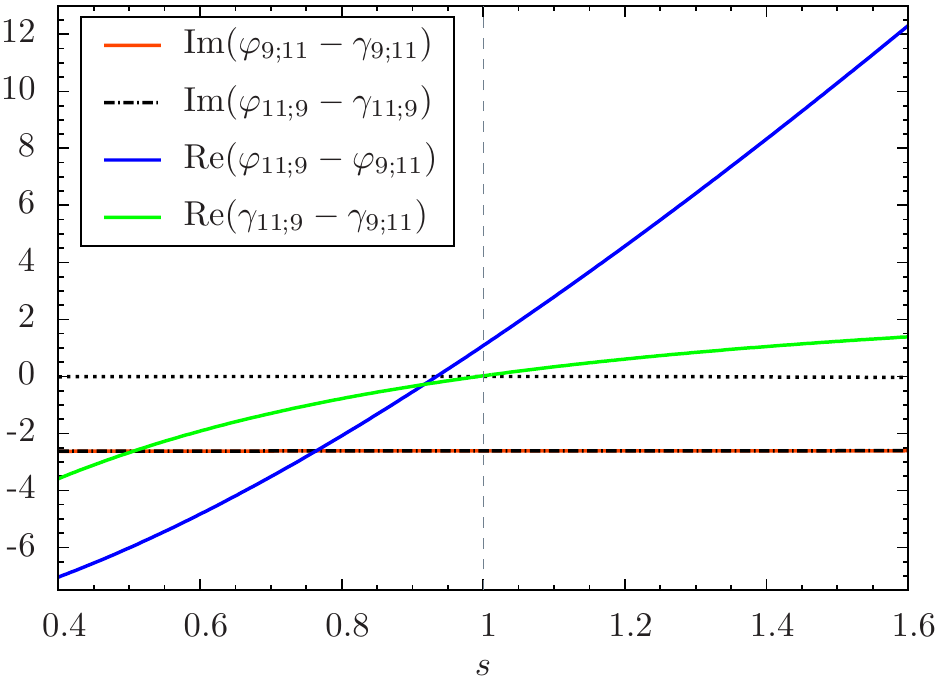}
\caption{The real and imaginary parts of all combinations of phase differences that can occur between the atomic states $\9$ and $\1$, using the approximate expression (\ref{Fpapprox}) for $\mathcal F_p(C_s)$. The black dotted line shows $-(\Delta\tau_\beta-\Delta\tau_\alpha)^2/(8\sigma'^2_k)$ for $C_s$ and $\beta=\9$ and $\alpha=\1$. For our realistic field configurations $\varphi_\9(z_{a})=\varphi_{\1}(z_{a})$ holds only approximately. The $s$-dependent deviations of the dynamic phases from the spin echo point $\varphi_\9(z_{a})=\varphi_{\1}(z_{a})$ lead to the specific interference patterns in Figures \ref{example1zeroE1SpinechoANDrev} and \ref{example1SpinechoANDrev}.}
\label{SpinEchoPuzzle20141121data}
\end{figure}

\begin{figure}[t]
\centering
\includegraphics[width=0.9\linewidth]{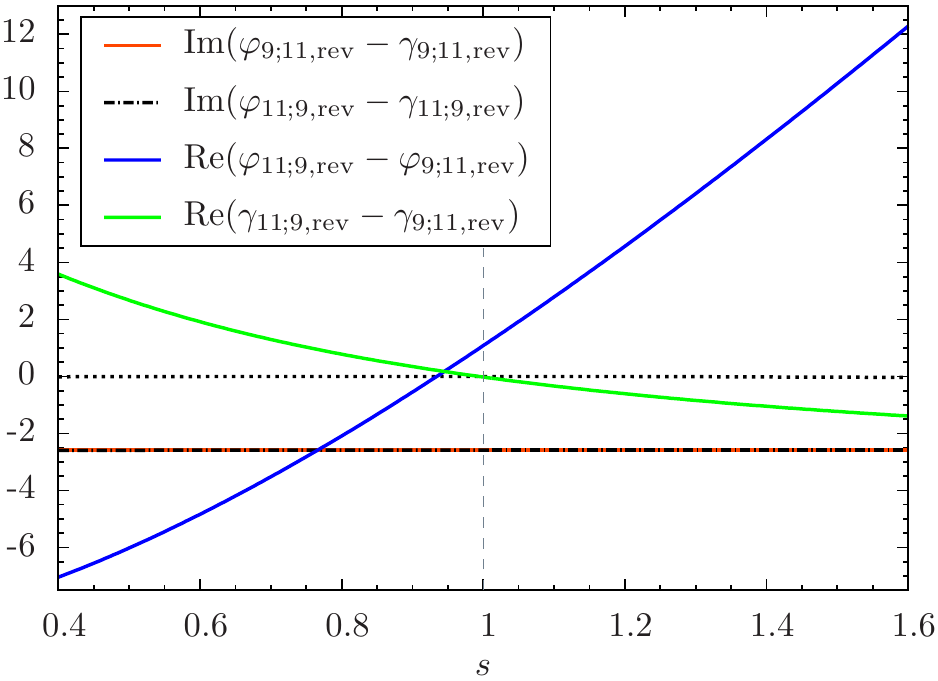}
\caption{The relevant quantities that compose the approximate expression (\ref{Fpapprox}) of $\mathcal F_p(\overline C_s)$.}
\label{SpinEchoPuzzle20141121reverseddata}
\end{figure}

As illustrated in Figure \ref{SpinEchoPuzzle20141121data2}, the approximation
\begin{align}
\exp\big[-(\Delta\tau_\beta-\Delta\tau_\alpha)^2/(8\sigma'^2_k)\big]\approx 1
\end{align}
holds at the percent level. 
Here $\Delta\tau_\alpha$ and $\sigma'_k$ are the shifts of the reduced arrival times and the momentum-space widths of the wave packets defined in (99) and (86) of Ref.~\cite{BeGaMaNaTr08_I}, respectively.
Furthermore,
\begin{align}
\label{eq45}
\mathrm{Im}\,\big(\varphi_{\9}-\gamma_{\9}\big)\approx \mathrm{Im}\,\big(\varphi_{\1}-\gamma_{\1}\big)
\end{align}
holds at the level of  per mille. Hence, the flux 
(\ref{Fp}) can be approximated by
\begin{align}\label{Fpapprox}
\mathcal F_p(C_s)\approx\ &\frac12\exp\big\{2\,\mathrm{Im}\,\big[\varphi_{\9}(z_a)-\gamma_{\9}(z_a)\big]\big\}\nn\\
\times\big(1&+\cos\big\{\mathrm{Re}\big[\varphi_{\1}(z_a)-\varphi_{\9}(z_a)\big]\nn\\
&\left.-\ \mathrm{Re}\big[\gamma_{\1}(z_a)-\gamma_{\9}(z_a)\big]\big\}\big)\right|_{C_s}
\end{align}
with a similar expression for $\mathcal F_p(\overline C_s)$,
\begin{align}\label{FpapproxBar}
\mathcal F_p(\overline C_s)\approx\ &\frac12\exp\big\{2\,\mathrm{Im}\,\big[\varphi_{\1}(z_a)-\gamma_{\1}(z_a)\big]\big\}\nn\\
\times\big(1&+\cos\big\{\mathrm{Re}\big[\varphi_{\1}(z_a)-\varphi_{\9}(z_a)\big]\nn\\
&\left.-\ \mathrm{Re}\big[\gamma_{\1}(z_a)-\gamma_{\9}(z_a)\big]\big\}\big)\right|_{\overline C_s}\,;
\end{align}
see Section 5.4 of \cite{DissMIT}. 
In (\ref{FpapproxBar}) we again make use of (\ref{eq45}) but we now write $\mathrm{Im}(\varphi-\gamma)$ with index $\1$ to recall the label change when going over from the curve $C_{s}$ to the reversed curve $\overline{C}_{s}$; see Figure \ref{spinflip}.
 The functions occuring in (\ref{Fpapprox}) and (\ref{FpapproxBar})  have been calculated for the realistic field configurations of Figure \ref{example1fields} and are shown in Figures \ref{SpinEchoPuzzle20141121data} and \ref{SpinEchoPuzzle20141121reverseddata} for $C_s$ and $\overline C_s$, respectively. 
The results are close to fulfilling the symmetry relations 
\begin{align}
&\left.\mathrm{Re}\big[\varphi_{\1}(z_a)-\varphi_{\9}(z_a)\big]\right|_{C_s}\nn\\
&\qquad=\left.\mathrm{Re}\big[\varphi_{\1}(z_a)-\varphi_{\9}(z_a)\big]\right|_{\overline C_s}\ ,\label{44a}\\
&\left.\mathrm{Re}\big[\gamma_{\1}(z_a)-\gamma_{\9}(z_a)\big]\right|_{C_s}\nn\\
&\qquad=-\left.\mathrm{Re}\big[\gamma_{\1}(z_a)-\gamma_{\9}(z_a)\big]\right|_{\overline C_s}\ .\label{44b}
\end{align}
In the ideal case where (\ref{symmcond}) holds we would also expect
\begin{align}
&\left.\mathrm{Re}\big[\varphi_{\1}(z_a)-\varphi_{\9}(z_a)\big]\right|_{C_1}\nn\\
&\qquad=\left.\mathrm{Re}\big[\varphi_{\1}(z_a)-\varphi_{\9}(z_a)\big]\right|_{\overline C_1}=0\ ,\label{44c}\\
&\left.\mathrm{Re}\big[\gamma_{\1}(z_a)-\gamma_{\9}(z_a)\big]\right|_{C_1}\nn\\
&\qquad=\left.\mathrm{Re}\big[\gamma_{\1}(z_a)-\gamma_{\9}(z_a)\big]\right|_{\overline C_1}=0\ .\label{44d}
\end{align}
%
%=====================================
\captionsetup[figure]{format=plain,font={scriptsize,normalfont}}
\renewcommand{\baselinestretch}{1.0}
\renewcommand{\baselinestretch}{\bls}
\clearcaptionsetup{figure}
%=====================================
%=====================================
\captionsetup[figure]{format=plain,font={scriptsize,normalfont}}
\renewcommand{\baselinestretch}{1.0}
\renewcommand{\baselinestretch}{\bls}
\clearcaptionsetup{figure}
%=====================================
%=====================================
\captionsetup[figure]{format=plain,font={scriptsize,normalfont}}
\renewcommand{\baselinestretch}{1.0}
\renewcommand{\baselinestretch}{\bls}
\clearcaptionsetup{figure}
%=====================================
We see from Figures \ref{SpinEchoPuzzle20141121data} and \ref{SpinEchoPuzzle20141121reverseddata} that for realistic fields, the symmetry relations (\ref{44a}), (\ref{44b}) and (\ref{44d}) are rather well satisfied, but (\ref{44c}) not so well. We shall now expand the relevant functions around $s=1$:
\begin{align}
&\left.\mathrm{Re}\big[\varphi_{\1}(z_a)-\varphi_{\9}(z_a)\big]\right|_{C_s}\nn\\
&\qquad=c_\varphi+m_\varphi\,(s-1)+\dots\ ,\\
&\left.\mathrm{Re}\big[\varphi_{\1}(z_a)-\varphi_{\9}(z_a)\big]\right|_{\overline C_s}\nn\\
&\qquad=\overline c_\varphi+\overline m_\varphi\,(s-1)+\dots\ ,\\
&\left.\mathrm{Re}\big[\gamma_{\1}(z_a)-\gamma_{\9}(z_a)\big]\right|_{C_s}\nn\\
&\qquad=c_\gamma+m_\gamma\,(s-1)+\dots\ ,\\
&\left.\mathrm{Re}\big[\gamma_{\1}(z_a)-\gamma_{\9}(z_a)\big]\right|_{\overline C_s}\nn\\
&\qquad=\overline c_\gamma+\overline m_\gamma\,(s-1)+\dots\ .
\end{align}
From Figures \ref{SpinEchoPuzzle20141121data} and \ref{SpinEchoPuzzle20141121reverseddata} we take that
\begin{align}\label{45e}
\begin{split}
c_\varphi&\approx\overline c_\varphi\ ,\\
m_\varphi&\approx\overline m_\varphi\ ,\\
m_\gamma&\approx-\overline m_\gamma\ ,\\
m_\varphi&>m_\gamma>0\ ,
\end{split}
\end{align}
and
\begin{align}\label{45ee}
\begin{split}
c_\gamma\approx\overline c_\gamma\approx 0\ .
\end{split}
\end{align}
Keeping only the constant terms and  those linear in $s-1$, which is a valid approximation when $|s-1|\lesssim0.3$, we can approximate the cosine in (\ref{Fpapprox}), for $C_s$, as
\begin{align}\label{cos1}
\cos\big[c_\varphi-c_\gamma+(m_\varphi-m_\gamma)(s-1)\big]
\end{align}
and in (\ref{FpapproxBar}), for $\overline C_s$, as
\begin{align}\label{cos2}
\cos\big[\overline c_\varphi-\overline c_\gamma+(\overline m_\varphi-\overline m_\gamma)(s-1)\big]\ .
\end{align}
Let us first consider $s=1$ for which we get, from (\ref{Fpapprox}),  (\ref{FpapproxBar}), (\ref{45e}) and (\ref{45ee}),
\begin{align}
\mathcal F_p(C_1)&\approx\left.\frac12\exp\big\{2\,\mathrm{Im}\,\big[\varphi_{\9}(z_a)-\gamma_{\9}(z_a)\big]\big\}\right|_{C_1}\nn\\
&\quad\times\big[1+\cos(c_\varphi)\big]\ ,\\
\mathcal F_p(\overline C_1)&\approx\left.\frac12\exp\big\{2\,\mathrm{Im}\,\big[\varphi_{\1}(z_a)-\gamma_{\1}(z_a)\big]\big\}\right|_{\overline C_1}\nn\\
&\quad\times\big[1+\cos(\overline c_\varphi)\big]\ ,
\end{align}
and for their ratio, using (\ref{example1Gammaeff}), (\ref{R}), (\ref{Ra}), and $c_\varphi\approx\overline c_\varphi$,
\begin{align}\label{R2}
\frac{\mathcal F_p(\overline C_1)}{\mathcal F_p(C_1)}
&\approx\ \frac{\exp[-T\,\Gamma_{\9,\mathrm{eff}}(\overline{C}_1)]}{\exp[-T\,\Gamma_{\9,\mathrm{eff}}(C_1)]}
\nonumber\\
&=\ \frac{\exp[-T\,\Gamma_{\1,\mathrm{eff}}(\overline{C}_1)]}{\exp[-T\,\Gamma_{\9,\mathrm{eff}}(C_1)]}
=1.057\ .
\end{align}
The discrepancy between this value for the quotient and 
\begin{align}\label{Rs1}
\frac{\mathcal F_p(\overline C_1)}{\mathcal F_p(C_1)}\approx\frac{4.1271}{4.0091}\approx1.0294\ ,
\end{align}
extracted from Figure \ref{example1SpinechoANDrevs1}, is due to the violation of (\ref{45ee}) by our realistic field configuration,
\begin{align}
c_\gamma\approx-\overline c_\gamma\approx0.022\ .
\end{align}
With $c_\varphi\approx\overline c_\varphi\approx1.079$ (see Figures \ref{SpinEchoPuzzle20141121data} and \ref{SpinEchoPuzzle20141121reverseddata}) we find, using (\ref{Fpapprox}), that
\begin{align}
\frac{\mathcal F_p(\overline C_1)}{\mathcal F_p(C_1)}&\approx\frac{\exp[-T\,\Gamma_{\1,\mathrm{eff}}(\overline{C}_1)]}{\exp[-T\,\Gamma_{\9,\mathrm{eff}}(C_1)]}\frac{1+\cos(\overline c_\varphi-\overline c_\gamma)}{1+\cos(c_\varphi-c_\gamma)}\nn\\
&\approx1.0295\ ,
\end{align}
consistent with (\ref{Rs1}). This observation underpins the necessity to measure the spin-echo curves for realistic field configurations over a sufficiently large $s$-range. In a followup experiment it will be necessary to make fits to $\mathcal F_p(C_s)$ and $\mathcal F_p(\overline C_s)$ and extract the imaginary parts of the geometric phases from these. We now show, that, \eg, the heights of the maxima of the spin-echo curves in Figure \ref{example1SpinechoANDrev} can be used for this purpose. With (\ref{45e})--(\ref{cos2}) we get, for $|s-1|\lesssim0.3$, the approximate expressions
\begin{align}
&\mathcal F_p(C_s)\approx\left.\frac12\exp\big\{2\,\mathrm{Im}\,\big[\varphi_{\9}(z_a)-\gamma_{\9}(z_a)\big]\big\}\right|_{C_s}\nn\\
&\times\big\{1+\cos\big[c_\varphi-c_\gamma+(m_\varphi-m_\gamma)(s-1)\big]\big\}\ ,\label{46c}\\
&\mathcal F_p(\overline C_s)\approx\left.\frac12\exp\big\{2\,\mathrm{Im}\,\big[\varphi_{\1}(z_a)-\gamma_{\1}(z_a)\big]\big\}\right|_{\overline C_s}\nn\\
&\times\big\{1+\cos\big[c_\varphi-\overline c_\gamma+(m_\varphi+m_\gamma)(s-1)\big]\big\}\ .\label{46d}
\end{align}
With $m_\varphi>m_\gamma>0$, see (\ref{45e}), we find that,  near $s=1$, $\mathcal F_p(\overline C_s)$ should oscillate with higher frequency than $\mathcal F_p(C_s)$. We see from Figure \ref{example1SpinechoANDrev} that this is indeed the case. At the maxima of $\mathcal F_p(C_s)$ and $\mathcal F_p(\overline C_s)$ that are the nearest to $s=1$ the cosines in (\ref{46c}) and (\ref{46d}) are equal to $1$ and the ratio of the fluxes is determined by the effective lifetimes of the states. We find these maxima for $\mathcal F_p(C_s)$ ($\mathcal F_p(\overline C_s)$) for $s=0.917$ ($s=0.945$) and $s=1.342$ ($s=1.253$). For the ratios of the fluxes we get
\begin{align}
\frac{\exp[-T\,\Gamma_{\1,\mathrm{eff}}(\overline{C}_1)]}{\exp[-T\,\Gamma_{\9,\mathrm{eff}}(C_1)]}\approx\frac{\mathcal F_p(\overline C_{0.945})}{\mathcal F_p(C_{0.917})}=1.060\ ,\label{R1}\\
\frac{\exp[-T\,\Gamma_{\1,\mathrm{eff}}(\overline{C}_1)]}{\exp[-T\,\Gamma_{\9,\mathrm{eff}}(C_1)]}\approx\frac{\mathcal F_p(\overline C_{1.253})}{\mathcal F_p(C_{1.342})}=1.051\ .\label{R1b}
\end{align}

As argued above, small uncertainties and asymmetries in a realistic experimental setup can lead to shifts and distortions of the spin-echo curves as compared to ideally symmetric field configurations. To extract the changes in atomic lifetimes at a desired confidence level, it is therefore in general not sufficient to measure the flux of atoms for a single value of $s$. We rather have to determine the spin-echo curves within a range of $s$ that includes the maxima of $\mathcal F_p(C_s)$ around $s=1$ and then invoke the same procedure for $\mathcal F_p(\overline C_s)$. Between $s=0.8$ and $s=1.4$ the two maxima of $\mathcal F_p(C_s)$ have approximately the same values, and the same holds for the maxima of $\mathcal F_p(\overline C_s)$. Therefore, both, the maxima for $s>1$ and $s<1$ serve to determine the geometry-induced relative changes in atomic lifetimes within the range $0.8\lesssim s\lesssim 1.4$. We can regard the difference between (\ref{R1}) and (\ref{R1b}) as a rough measure of the uncertainty of our prediction for the geometric lifetime effects given the imperfections of a realistic field configuration. For other field configurations the quantities entering in $\mathcal F_p$ have to be investigated analogously to determine whether the lifetime modifications can be extracted from the maxima of the spin-echo curves.

%=====================================
\captionsetup[figure]{format=plain,font={scriptsize,normalfont}}
\renewcommand{\baselinestretch}{1.0}
\begin{figure}[t]
\centering
\includegraphics[width=0.99\linewidth]{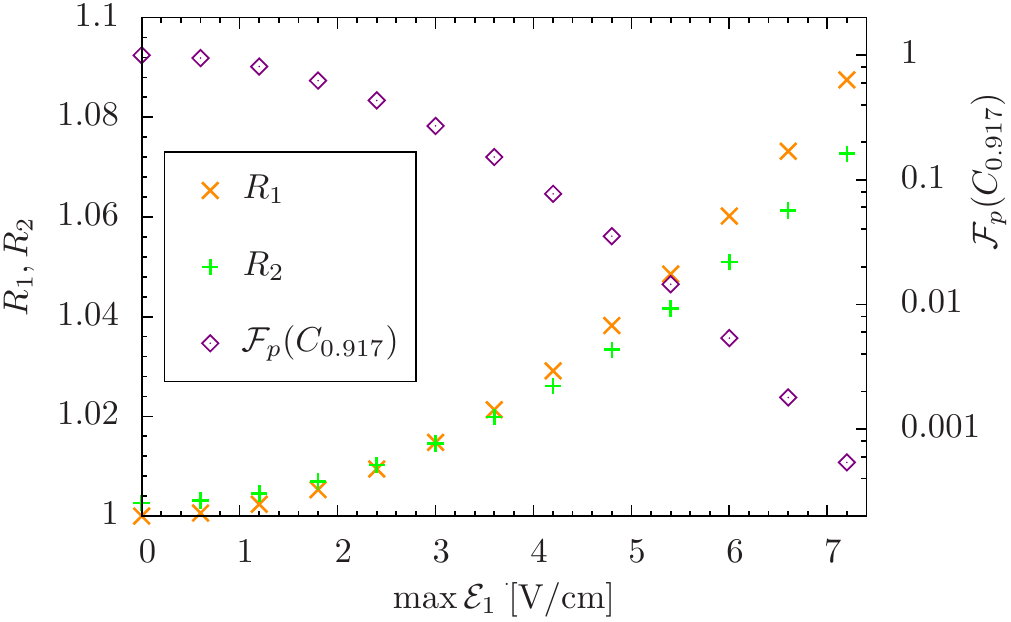}
\caption{The ratios $R_1=\mathcal F_p(\overline C_{0.945})/\mathcal F_p(C_{0.917})$ and $R_2=\mathcal F_p(\overline C_{1.253})/\mathcal F_p(C_{1.342})$ as well as the flux $\mathcal F_p(C_{0.917})$ as a function of max$\,\mathcal E_1$. The electric field in Figure \ref{example1fields} corresponds to max$\,\mathcal E_1=6\,$V/cm. At max$\,\mathcal E_1=0$ we find $R_1\approx1$. There, the corresponding maxima $\mathcal F_p(\overline C_{0.945})$ and $\mathcal F_p(C_{0.917})$ of the spin-echo curves, separated by $\Delta s\approx0.028$ and relatively close to $s=1$, are practically of the same height. This is not the case for $R_2$ due to the fact that the maxima are found at $s$-values which differ significantly more, $\Delta s\approx0.089$; cf.~Figure \ref{example1zeroE1SpinechoANDrev}. We see here that comparison of fluxes at different values of $s$ that are separated by large $\Delta s$ can increasingly mask the effect of the geometric phases on the decay rates. At which values of $s$ we can find the best estimate of the geometry-induced lifetime modification depends on the position of the maxima relative to $s=1$.}
\label{RampUpE1}
\end{figure}
\renewcommand{\baselinestretch}{\bls}
\clearcaptionsetup{figure}
%=====================================

We now study the dependence of the geometric lifetime effects on the applied electric field component $\mathcal{E}_{1}$.
In the range where $\mathcal{E}_{1,\mathrm{max}}\le7.2\,$V/cm we find that the maxima of the spin-echo curves, both for $\mathcal{F}_{p}(C_{s})$ and $\mathcal{F}_{p}(\overline C_{s})$, remain essentially at the same values of $s$ as extracted from Figure \ref{example1SpinechoANDrev}. 
We therefore show, in Figure \ref{RampUpE1}, the ratios
\begin{align}
R_1=\frac{\mathcal F_p(\overline C_{0.945})}{\mathcal F_p(C_{0.917})}
\end{align}
and
\begin{align}
R_2=\frac{\mathcal F_p(\overline C_{1.253})}{\mathcal F_p(C_{1.342})}
\end{align}
at the same values of $s$ as in (\ref{R1}) and (\ref{R1b}), respectively.
We also show the absolute value of the spin-echo signal $\mathcal F_p(C_{0.917})$ as a function of the magnitude of $\mathcal E_1$. The field $\mathcal E_1$ is scaled such that its maximum ranges between $0$ and $7.2\,$V/cm. The ratio $R_1$ increases also for electric fields larger than $\mathcal E_{1,\mathrm{max}}=6\,$V/cm, but at the expense of the count rate which is proportional to $\mathcal F_p(C_{0.917})$. We chose $\mathcal E_{1,\mathrm{max}}=6\,$V/cm, see Figure \ref{example1fields}, as a reasonable compromise between the observable relative effect on the lifetimes and experimental feasibility.

The measurement of $\mathcal F_p$ can be considered as measurement of a random variable $\xi$ taking on two values.
We set $\xi=1$ if an atom is detected at $z=z_{a}$ and $\xi=0$ if no atom is detected.
In the latter case the atom may have decayed before arriving at $z_{a}$ or it may be in a state orthogonal to the analysing state $\rket{p}$ at $z_{a}$; cf.~(\ref{palpha}).
Suppose now that we start with one atom at $z=0$.
Then the probability to get $\xi=1$ is given by $\mathcal F_p$, the probability to get $\xi=0$ is $1-\mathcal F_p$.
Thus, we have for the expectation value and the variance of $\xi$
\begin{align}
E_{1}(\xi) &= \mathcal F_p\,,\nn\\
\mathrm{Var}_{1}(\xi) &= E_{1}(\xi^{2})-[E_{1}(\xi)]^{2}\nn\\
	&=\mathcal F_p-\mathcal F_p^{2}=\mathcal F_p(1-\mathcal F_p)\, . 
\end{align}
Next we suppose that we start with $N$ atoms.
Then we get for the average $\bar\xi$ the following expectation value and variance:
\begin{align}
E_{N}(\bar\xi) &= \mathcal F_p\,,\nn\\
\mathrm{Var}_{N}(\bar\xi) &= \frac1N\mathrm{Var}_{1}(\xi)= \frac1N\mathcal F_p(1-\mathcal F_p)\, . 
\end{align}
If we want to measure $\mathcal F_p$ with a relative accuracy $\delta$ we should achieve
\begin{align}
[\mathrm{Var}_{N}(\bar\xi)]^{1/2} <\delta E_{N}(\bar\xi)\, , 
\end{align}
that is,
\begin{align}
\frac1N\mathcal F_p(1-\mathcal F_p)<\delta^{2}\mathcal F_p^{2}\, . 
\end{align}
This requires the number of atoms to obey
\begin{align}
N>\frac{1}{\delta^{2}}\left(\frac{1}{\mathcal F_p}-1\right)\, . 
\label{Nmindelta}
\end{align}
We consider, as a representative value of $\mathcal F_p$, half the maximum value $\mathcal F_p(C_{0.917})$ at $\mathrm{max}\,\mathcal E_{1}=6\,$V/cm.
From Figure \ref{RampUpE1} we then find $\mathcal F_p(C_{0.917})/2\approx2.7\times10^{-3}$.
For a $0.5\%$ measurement of this value of $\mathcal F_p$, condition (\ref{Nmindelta}) requires us to work with 
\begin{align}
N>1.5\times10^{7}\,  
\end{align}
atoms to obtain one data point on the spin-echo curve.
To measure the complete spin-echo curves we will demand $100$ data points for each, $C_{s}$ and $\overline C_{s}$.
Hence, the total number of atoms needed is $N>3\times10^{9}$.
With the corresponding accuracy of $0.5\%$ per data point of $\mathcal F_{p}$ on the spin-echo curves\footnote{%
The theoretical error of $\mathcal F_{p}$ is estimated to be of the same order; see \cite{BeGaMaNaTr08_I}.}
it should be possible to obtain an accuracy of $10\%$ for our geometric lifetime effect which is of the order of $5$ to $6\%$.

%==========================================================================
%==========================================================================
\subsection{Conclusions}

In this article we calculate the lifetime modification of metastable states of hydrogen due to geometric phases. A geometry-induced modification of atomic decay rates has not been observed experimentally thus far. In addition to imaginary \textit{dynamic} phases, which emerge in an effective description of decaying atomic states travelling in an adiabatic way through electromagnetic fields, the hydrogen state vectors acquire imaginary \textit{geometric} phases in suitable chiral electromagnetic field configurations. We use the time evolution of a superposition of metastable states propagating in a field configuration which is based on realistic experimental conditions to compute the flux of atoms arriving at the detector of a longitudinal atomic-beam spin-echo apparatus. We analyse the relevant quantities entering the description of the propagating atomic wave packet, in particular the dynamic and geometric phases, and propose a realistic scheme to observe the change of lifetimes experimentally. We ensure adiabatic evolution in spatial regions where geometric phases for the hydrogen state vectors emerge. We vary the field configuration to obtain spin-echo curves which are conveniently accessible in experiment. We show in detail how to extract the geometry-induced change of lifetime from the maxima of the spin-echo curves and estimate the necessary number of metastable atoms to be $4\times10^{9}$ for a statistically significant measurement. We find that the lifetime is modified at the level of $5\%$ due to geometric phases. We estimate that this effect is large enough to be observed under realistic experimental conditions.
\newpage

%==========================================================================
%==========================================================================
%==========================================================================
%==========================================================================
\appendix

\renewcommand{\thesubsection}{\Alph{subsection}}
\renewcommand{\theequation}{\thesubsection.\arabic{equation}}  

\section*{Appendix}
%==========================================================================
%==========================================================================
\subsection{Conditions for adiabatic evolution of the states}\label{AppendixAdiabat}
\setcounter{equation}{0}

Employing the field configuration from Figure \ref{example1fields} with $s=1$, we find that the adiabaticity conditions (B.16) and (B.22) from \cite{BeGaNa07_II} for the field variations are satisfied. We get
\begin{align}
\frac{1}{\mathcal E_0}\underset{t\in [0,T]}{\mathrm{max}}\left|\frac{\partial\calEvec}{\partial t}\right|&<\frac{1}{T}\longleftrightarrow 0.69\lesssim 1\,,\\
\frac{1}{\mathcal B_0}\underset{t\in[0,T]}{\mathrm{max}}\left|\frac{\partial\calBvec}{\partial t}\right|&<\frac{1}{T}\longleftrightarrow 0.1\lesssim 1\,, 
\end{align}
with $\mathcal E_0=477.3\,$V/cm, $\mathcal B_0=43.65\,$mT, $T=(z_a-z_0)/v_z$. Wherever geometric phases emerge along the $z$-axis the energy separation $\Delta E$ between the involved states is large enough for adiabatic evolution:
\begin{align}\label{adiabE}
\Delta z\gtrsim\frac{h\,v_z}{\Delta E}\longleftrightarrow 90\,\mathrm{mm}\gg 19\,\mathrm{mm}\ ; 
\end{align}
see (27) from \cite{BeGaMaNaTr08_I}. For $\Delta z$ we have $90\,$mm from the fields of Figure \ref{example1fields}. Of course, for $s>1$ the adiabaticity condition (\ref{adiabE}) is satisfied as well, whereas $s$ may not be taken much smaller than $s=1$.

%==========================================================================
%==========================================================================
\subsection{Field configuration}
\label{AppendixField}
\setcounter{equation}{0}

We employ a field configuration as depicted in Figure \ref{example1fields} which is actually available in the laboratory. The magnetic field components are fits to measured data. The electric field component $\mathcal E_1(z)$ is calculated via a finite-elements method and is experimentally realisable with an appropriate set of capacitor plates. It is straightforward to adjust the analysis presented in this work for slightly different experimental realisations of $\mathcal E_1(z)$. The remaining field components are chosen to be zero, the electric field is given in units of V/cm, the magnetic field components in units of $\mu$T. For the calculation of $\mathcal F_p(C_s)$ and $\mathcal F_p(\overline C_s)$ with $C_1$ illustrated in Figure \ref{example1fields}, we vary $s$ in the $z$-intervals $[0.32330919,0.66]$ and $[0,0.33669081]$, respectively. We define
\begin{align}
\begin{split}
z_m&=0.33380917\ ,\\
z_1&=0.31676777\ ,\\
z_2&=0.57813638\ .
\end{split}
\end{align}
Using
$c_0= -15625$, $c_1 = 0.009$, $c_2 = 0.0105$, $c_3 = 0.07$, $c_4 = 0.08$, $c_5 = 0.16$, $c_6 = 0.17$, and employing the syntax `A ? B : C' for `B to be true if A is, and C to be true if A is not' and use logical `AND' and `OR', the fields are given as
%\newpage
%
\newcommand{\storedisplaystyle}{\displaystyle}
\renewcommand{\displaystyle}{\scriptstyle}

\begin{widetext}
%\onecolumngrid
%\vspace{\columnsep}
%
\begin{align}
\mathcal{E}_1(z)\,=\,& 3\,\big\{\left(z + c_1 > z_m - c_5 \;\mathrm{AND}\; z + c_1 < z_m - c_3\right) \;\mathrm{OR}\; \left(z + c_1 > z_m + c_4 \;\mathrm{AND}\; z + c_1 < z_m + 0.17\right) \;\,?\;\, 1\;\, :\;\,\nn\\
&\exp \left[c_0\,\left(z + c_1 - (z_m - c_5)\right)\,\left(z + c_1 - (z_m - c_5)\right)\right] + \exp \left[c_0\,\left(z + c_1 - (z_m - c_3)\right)\,\left(z + c_1 - (z_m - c_3)\right)\right]\nn\\
&+\exp \left[c_0\,\left(z + c_1 - (z_m + c_4)\right)\,\left(z + c_1 - (z_m + c_4)\right)\right] + \exp \left[c_0\,\left(z + c_1 - (z_m + c_6)\right)\,\left(z + c_1 - (z_m + c_6)\right)\right]\nn\\
&+ \left(z_a + c_1 - z > z_m - c_5 \;\mathrm{AND}\; z_a + c_1 - z < z_m - c_3\right) \;\mathrm{OR}\; \left(z_a + c_1 - z > z_m + c_4 \;\mathrm{AND}\; z_a + c_1 - z < z_m + c_6\right) \;\,?\;\, 1 \;\, :\;\,\nn\\
&\exp \left[c_0\,\left(z_a + c_1 - z - (z_m - c_5)\right)\,\left(z_a + c_1 - z - (z_m - c_5)\right)\right] + \exp \left[c_0\,\left(z_a + c_1 - z - (z_m - c_3)\right)\,\left(z_a + c_1 - z - (z_m - c_3)\right)\right]\nn\\
&+\exp \left[c_0\,\left(z_a + c_1 - z - (z_m + c_4)\right)\,\left(z_a + c_1 - z - (z_m + c_4)\right)\right] + \exp \left[c_0\,\left(z_a + c_1 - z - (z_m + c_6)\right)\,\left(z_a + c_1 - z - (z_m + c_6)\right)\right]\big\}\\
\mathcal{B}_1(z)\,=\,& z + c_2 < z_m\;\,?\;\,\big\{z + c_2 < z_1\;\,?\;\,-153.283\,\exp [-2251.75\,(z + c_2 - 0.221739)^2]\, \sin [3.95282\,(z + c_2 - 0.222097)]\nn\\
&-50.294\,\exp [-2174.46\,(z + c_2 - 0.457696)^2]\, \sin [11.8633\,(z + c_2 - 0.193543)] \;\, :\;\,\nn\\
&\big[z + c_2 < 0.36223008\;\,?\;\, 0 \;\, :\;\, \{z + c_2 < z_2\;\,?\;\,-153.283\,\exp [-2251.75\,(z + c_2 - 0.221739)^2]\,\sin [3.95282\,(z + c_2 - 0.222097)]\nn\\
&-50.294\,\exp [-2174.46\,(z + c_2 - 0.457696)^2]\,\sin [11.8633\,(z + c_2 - 0.193543)] \;\, :\;\, 0\}\big]\big\} \;\, :\;\,\nn\\
&\big\{z + c_2 < z_1\;\,?\;\,153.283\,\exp [-2251.75\,(z + c_2 - 0.221739)^2]\,\sin [3.95282\,(z + c_2 - 0.222097)]\nn\\
&50.294\,\exp [-2174.46\,(z + c_2 - 0.457696)^2]\,\sin [11.8633\,(z + c_2 - 0.193543)] \;\, :\;\,\nn\\
&\big[z + c_2 < 0.36223008\;\,?\;\, 0 \;\, :\;\, \{z + c_2 < z_2\;\,?\;\,153.283\,\exp [-2251.75\,(z + c_2 - 0.221739)^2]\,\sin [3.95282\,(z + c_2 - 0.222097)]\nn\\
&50.294\,\exp [-2174.46\,(z + c_2 - 0.457696)^2]\,\sin [11.8633\,(z + c_2 - 0.193543)] \;\, :\;\, 0\}\big]\big\}\\
\mathcal{B}_2(z)\,=\,& z + c_2 < z_1\;\,?\;\,\big\{32.34\, \exp [-490.685\,(z + c_2 - 0.222096)^2]\,(\cos^2[29.8529\,(z + c_2 - 0.222682)] - 0.894862)\nn\\
&-34.967\, \exp [-515.945\,(z + c_2 - 0.458747)^2]\,(\cos^2[29.0659\,(z + c_2 - 0.459074)] - 0.901612)\big\} \;\, :\;\, \big\{z + c_2 < 0.36223007941\;\,?\;\, 0 \;\, :\;\,\nn\\
&\big(z + c_2 < z_2\;\,?\;\,32.34\,\exp [-490.685\,(z + c_2 - 0.222096)^2]\,(\cos^2[29.8529\,(z + c_2 - 0.222682)] - 0.894862)\nn\\
&-34.967\,\exp [-515.945\,(z + c_2 - 0.458747)^2]\,(\cos^2[29.0659\,(z + c_2 - 0.459074)] - 0.901612) \;\, :\;\, 0\big)\big\}\\
\mathcal{B}_3(1;z)\,=\,&1.4\times10^{-3} - 7.476112\,\exp [-535.705\,(z + c_2 - 0.451987)^2] + 7.482594\,\exp [-566.72\,(z + c_2 - 0.215842)^2]
\end{align}
%
%\vspace{\columnsep}
%\twocolumngrid
\end{widetext}
%\flushend
%\balance
\ \vspace*{-1.7cm}\\
%\clearpage
\renewcommand{\displaystyle}{\storedisplaystyle}

\bibliographystyle{mit_PhDThesis_BibitStyleFile} % manually produced with Bibit (to modify the style file (mit_PhDThesis_BibitStyleFile.bst) type in console (in bibit-folder): ./Bibit1_4.jar): -> Tools -> Style Generator -> choose 'mit_PhDThesis_BibitStyleFile.save' (enlarge the window to see all buttons...) -> modify things -> Generate and Save as mit_PhDThesis_BibitStyleFile.bst
%\bibliography{myapvbib_20130722}

\balance

\end{document}